\documentclass[journal]{IEEEtran}

\usepackage{cite}

\usepackage{color}
\usepackage{booktabs}
\usepackage{amsmath}

\usepackage{bm}
\usepackage{latexsym}
\usepackage{amsthm}
\usepackage{url}
\usepackage{amsfonts}
\usepackage{amssymb}
\usepackage{indentfirst}
\usepackage{float}
\usepackage{tabularray}
\usepackage[ruled,linesnumbered]{algorithm2e}

\ifCLASSINFOpdf
   \usepackage[pdftex]{graphicx}
   \graphicspath{{../pdf/}{../jpeg/}}
   \DeclareGraphicsExtensions{.pdf,.jpeg,.png}
\else
   \usepackage[dvips]{graphicx}
   \graphicspath{{../eps/}}
   \DeclareGraphicsExtensions{.eps}
\fi

\usepackage{array}
\usepackage{multirow}

\ifCLASSOPTIONcompsoc
  \usepackage[caption=false,font=normalsize,labelfont=sf,textfont=sf]{subfig}
\else
  \usepackage[caption=false,font=footnotesize]{subfig}
\fi

\usepackage{setspace}
\begin{document}

\title{ {Heterogeneous window Transformer for image denoising}}

\author{Chunwei Tian, \emph{Member}, \emph{IEEE},
        Menghua Zheng,
        Chia-Wen Lin, \emph{Fellow}, \emph{IEEE},
        Zhiwu Li, \emph{Fellow}, \emph{IEEE},
        David Zhang, \emph{Life Fellow}, \emph{IEEE}
        
\thanks{This work was supported in part by the National Natural Science Foundation of China under Grant 62201468, in part by the TCL Science and Technology Innovation Fund under Grant D5140240118. (Corresponding author: Chia-Wen Lin (cwlin@ee.nthu.edu.tw))}
\thanks{Chunwei Tian is with School of Software, Northwestern Polytechnical University and also with National Engineering Laboratory for Integrated Aero-Space-Ground-Ocean Big Data Application Technology, Xi’an, 710129, China. (email: chunweitian@nwpu.edu.cn)}
\thanks{Menghua Zheng is with School of Software, Northwestern Polytechnical University, Xi’an, 710129, China. (email:menghuazheng@mail.nwpu.edu.cn)}
\thanks{Chia-Wen Lin is with Department of Electrical Engineering and the Institute of Communications Engineering, National Tsing Hua University, 300, Taiwan. 
 (email:cwlin@ee.nthu.edu.tw)}
\thanks{Zhiwu Li is with the School of Elector-Mechanical Engineering, Xidian University, Xi'an, 710071, China. 
 (email:zhwli@xidian.edu.cn)}
\thanks{David Zhang is with School of Data Science, Chinese University of Hong Kong (Shenzhen), Shenzhen, 518172, China and also with Shenzhen Institute of Artificial Intelligence and Robotics for Society, Shenzhen, 518172, China.  (email:davidzhang@cuhk.edu.cn)}
}


\maketitle

\begin{abstract}
\textcolor{black}{Deep networks can usually depend on extracting more structural information to improve denoising results. However, they may ignore correlation between pixels from an image to pursue better denoising performance. Window Transformer can use long- and short-distance modeling to interact pixels to address mentioned problem. To make a tradeoff between distance modeling and denoising time,} we propose a heterogeneous window Transformer (HWformer) for image denoising. HWformer first designs heterogeneous global windows to capture global context information for improving denoising effects. To build a bridge between long and short-distance modeling, global windows are horizontally and vertically shifted to facilitate diversified information without increasing denoising time. To prevent the information loss phenomenon of independent patches, sparse idea is guided a feed-forward network to extract local information of neighboring patches. The proposed HWformer only takes 30\% of popular Restormer in terms of denoising time. 
 Its codes can be obtained at https://github.com/hellloxiaotian/HWformer.
\end{abstract}


\section{Introduction}
Image denoising techniques are dedicated to recovering clean images from given noisy images. That is, they depend on a degradation model of  $y=x+n$, where $y$ and $x$ denote a given noisy image and clean image, respectively. Also, $n$ is used to express noise, which is regarded to additive white Gaussian noise (AWGN) \cite{levin2011natural}.  Numerous traditional machine learning techniques can use the degradation model to obtain resolutions of the ill-posed problem for image denoising. Specifically, they usually use prior knowledge, i.e., sparse \cite{li2012group}, total variation \cite{beck2009fast} and non-local similarity \cite{maggioni2012nonlocal}  to suppress noise.  Although these methods have obtained good denoising results, they suffered from challenges of manual tuning parameters and complex optimization functions. 

Convolutional neural networks (CNNs) can obtain strong learning abilities via stacking  simple components, i.e., convolutional layers and activation functions to overcome mentioned challenges for video and image applications \cite{muhammad2018efficient, yu2020two}, especially image denoising \cite{zhang2017beyond}. Denoising CNN (DnCNN) uses convolutional layers, batch normalization techniques, rectified linear unit (ReLU) and batch normalization techniques to make a tradeoff between denoising performance and efficiency \cite{zhang2017beyond}. To promote denoising effects, residual dense network (RDN) integrated hierarchical information to facilitate richer detailed information for recovering clean images \cite{zhang2020residual}. Moreover, asymmetric CNN (ACNet)\cite{tian2021asymmetric} embedded asymmetric ideas into a CNN to enhance local features to restore more details for image resotration. Due to small kernels, CNNs can only obtain local information, which may limit denoising performance. Transformer uses pixel relations to capture global information to overcome drawbacks of CNNs for image denoising \cite{chen2021pre}. Image process Transformer (IPT) utilizes a Transformer containing a self-attention mechanism and a feed-forward network to interact global pixels for promoting denoising effects \cite{chen2021pre}. Alternatively, SwinIR restricts effect area of a self-attention mechanism to a local window to reduce computational cost for image denoising \cite{liang2021swinir}. 

Although the window Transformer can rely on a short-distance modeling to reduce denoising time, local windows can limit interactions of contexts. Also, existing Transformer based long-distance modeling will increase denoising time. To build a bridge between distance modeling and denoising time, we present a heterogeneous window Transformer (HWformer) in image denoising. HWformer first designs heterogeneous global windows to try best to guarantee interactions of self-attention mechanisms to obtain more global context information for promoting performance of image denoising. To break the limitation of interactions of long and short-distance modeling, global windows are horizontally and vertically shifted to facilitate diversified information without increasing denoising time. To prevent native effects of independent patches, sparse idea is first embedded into a feed-forward network to extract more local information of neighboring patches. Also, our HWformer only takes 30\% of popular Restormer in terms of denoising time.

Contributions of this paper can be summarized as follows. 

1. \textcolor{black}{Heterogeneous global windows with different sizes are designed to facilitate richer global context information to overcome drawback of short-distance modeling.} 

2. \textcolor{black}{A shift mechanism of different directions is first deigned in the global windows to build a bridge  between short and long-distance modeling to improve denoising performance without increasing denoising time.} 

3.	\textcolor{black}{Sparse technique is proposed in a feedforward network to capture more local 
information of neighboring patches in image denoising.}  

4.	Our HWformer has faster denoising speed, which has near three times of popular Restormer in image denoising.

Remaining parts of this paper are as follows. Section II provides related work of deep learning techniques for image denoising. Section III lists proposed work containing network architecture, loss function, global-window Transformer enhancement block and Transformer direction enhancement block. Section IV gives experimental analysis and results. Section V summaries the whole paper. 
\section{Related work}
\textcolor{black}{Although convolutional neural networks have powerful feature extraction abilities, they are still faced with challenges of data scarcity and imbalance for image denoising. In terms of data scarcity, data augmentation techniques are good choice for image denoising \cite{kim2019grdn}. Due to strong generative abilities, generative adversarial networks (GANs) are used for data augmentation to improve performance of image denoising \cite{yang2018low}. To increase the number of training data, Wasserstein
is embedded into a GAN to improve denoising effect \cite{luo2021gpr}. Yang et al.\cite{yang2018low} used optimal transport idea to enhance a GAN to enhance data for promoting denoising effect. To address real noisy image denoising, a two-step denoising method is presented \cite{tran2020gan}. That is, the first step maps given images as raw images. The second step uses a GAN to estimate noise distribution on a large scale collected images, where obtained noise distribution can be used to train a denoising model. Alternatively, Hong et al. \cite{hong2020end} utilized 
conditional generative adversarial network to learn noise distribution from given noisy images to achieve a image blind denoising. Fuentes et al. \cite{fuentes2022mid3a} fused a regularization term into a GAN to learn noise distribution and used structure preserving loss to improve denoising results. In terms of data imbalance, normalization techniques \cite{ioffe2015batch} are effective tools to address nonuniform data distribution in image denoising. For instance, Zhang et al. applied  batch normalization techniques, a residual learning operation to act a convolutional network to achieve an efficient denoising model, where batch normalization technique is used to normalize obtained structural information to make a tradeoff between training efficiency and denoising performance. 
To address non-uniformly distributed data affected by constrained-resource hardware platform, Tian et al. \cite{tian2020image} used batch renormalization technique to normalize whole sample rather than normalizing batch of batch normalization to improve denoising effect. To prevent overfitting, adaptive instance normalization is  gathered into a CNN to establish a denoiser \cite{kim2020transfer}. Mentioned illustrations can use structural information to improve performance of image denoising. However, pixel relations can improve effects of image denoising. Thus,  we combine 
 structural relation and pixel relation to improve denosing results in this paper.} 
\begin{figure*}[h]
  \centering
  \includegraphics[width=\linewidth]{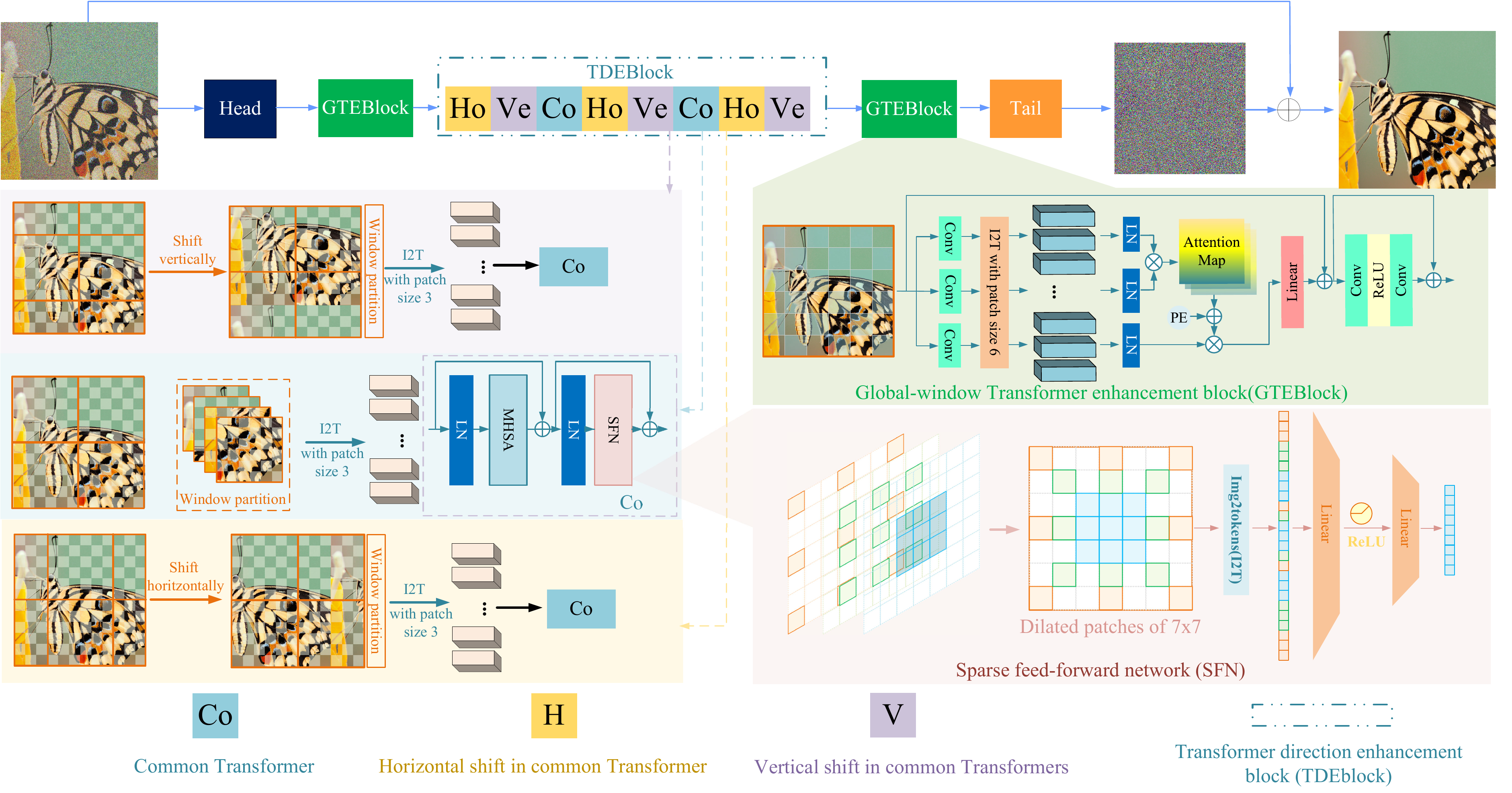}
  \caption{Network architecture of HWformer.}
  
\end{figure*}

\section{Proposed work}

In this section, we introduce the overall architecture of HWformer and its key techniques, i.e., stacked convolutional layers, global-window Transformer enhancement block (GTEBlocks) and a Transformer direction enhancement block (TDEBlock) in Fig.1. 

\subsection{Network architecture of HWformer}
HWformer composed of a head, two GTEBlocks, TDEBlock and a tail can be used to break a limitation of long- and short-distance modeling to improve denoising effects. That is, a head \cite{chen2021pre}  containing 5-layer convolutional layers, ReLU and residual learning operations is used to extract shallow information. To capture richer global information, we design a heterogeneous architecture containing two GTEBlocks and a TDEBlock. Specifically, each GTEBlock enlarges input windows to extract coarse global information. Also, GTEBlock gathers convolutional layers into a Transformer to facilitate richer information, i.e., structural information and information of pixel relations, which can overcome challenges of short-distance modeling for image denoising. Taking into merits of short-distance modeling in denoising speed account, TDEBlock is designed. TDEBlock is composed of eight Transformers to extract fine global information. Each Transformer is one of three kinds, horizontal (also regarded as Ho), vertical (also regarded as Ve) and common (also regarded as Co) window Transformer mechanisms. It is also a heterogeneous, which can facilitate richer information for image denoising. Besides, we first use sparse idea into Ho, Ve and Co to extract more local information from neighboring patches. \textcolor{black}{Also, pseudocode of Ho, Ve, Co can be shown in Algorithm 1.} A tail is composed of a convolutional layer, which is used to construct high-quality images. 

\subsection{Loss function}

\textcolor{black}{
To train a more robust model and facilitate a fair comparison, we choose popular mean square error (MSE) \cite{allen1971mean} as loss function to optimize the parameters of HWformer. The loss function can be represented as follows:
}

\begin{equation}
\begin{array}{ll}
L(\theta ) = \frac{1}{{2N}}\sum\limits_{i = 1}^N {\left\| {HWformer(I_n^i) - I_c^i} \right\|^2} 
\end{array}
\end{equation}

where $I_n^i$ and $I_c^i$ stand for the paired $ith$ noisy and clean images. Also, $N$ represents the number of noisy images in the training dataset. And the total parameters $\theta$ of the model can be optimized via Adam\cite{kingma2014adam}.

\subsection{Global-window Transformer enhancement block}

To break the limits of short-distance modeling, GTEBlock uses global window rather than local window to extract global information. That is, GTEBlock cuts into windows of $96\times96$  rather than that of  $48\times48$ \cite{chen2021pre} in the self-attention to enlarge receptive field for capturing more global information, which can also break the limits of global information loss of short-distance modeling. That is, obtained 2D features input into three independent convolutional layers rather than fully connected layers to obtain query (Q), key (K) and value (V), where used convolutional layers can reduce parameters. Then, GTEBlock cuts respectively Q, K and V to non-overlapping patches of $p\times p$ to reduce computational costs and they are flattened as vectors. Next, to make training stable, obtained vectors are normalized \cite{ba2016layer} as inputs of a multi-head self-attention mechanism to achieve global interactions of pixels, where the multi-head self-attention mechanism can be obtained at Ref. \cite{liang2021swinir}. Finally, a combination of two convolutional layers and ReLU is rather than 2-layer feedforward network to reduce the number of parameters. To prevent long-term dependency problem, two residual learning operations are applied in each GTEBlock as shown in Fig.1. \textcolor{black}{Additionally, its pseudocode can be shown in Algorithm 2.} 

\begin{table*}
\centering
  \caption{PSNR of different models for image denoising on Urban100\cite{huang2015single} with noise level of 15.}
  \label{tab:freq}
  \begin{tabular}{ccccccccc}
    \toprule
    IDs      & Co(number) & Ho(number)     & Ve(number) & GTEBlock(number) & FN & SFN & PSNR  &   \\
    \midrule
1       &\checkmark(10)     & ~              & ~          & ~       &\checkmark         & ~      & 33.72 & ~ \\
2       &\checkmark(5)      &\checkmark(5)          & ~          & ~    &\checkmark             & ~      & 33.80 & ~ \\
3       &\checkmark(5)      & ~              &\checkmark(5)      & ~     &\checkmark           & ~      & 33.79 & ~ \\
4       &\checkmark(4)      &\checkmark(3)          &\checkmark(3)      & ~      &\checkmark          & ~      & 33.84 & ~ \\
5      &\checkmark(2)      &\checkmark(3)\textbf{} &\checkmark(3)      &\checkmark(2) with FCL  &\checkmark & ~      & 33.91 & ~ \\
6       &\checkmark(2)      &\checkmark(3)          &\checkmark(3)      &\checkmark(2)           &\checkmark & ~      & 33.91 & ~ \\
7       & ~          & ~              & ~          &\checkmark(10)      &~     & ~      & 33.80 & ~ \\
8       &\checkmark(2)      &\checkmark(3)\textbf{} &\checkmark(3)      &\checkmark(2)      &~      & ~      & 33.47 & ~ \\
9 (Ours) &\checkmark(2)      &\checkmark(3)          &\checkmark(3)      &\checkmark(2)      &~      &\checkmark     & 33.94 & ~ \\
  \bottomrule
\end{tabular}
\end{table*}

\begin{table*}[]
\centering
\caption{\textcolor{black}{PSNR of different models with different window size for image denoising on BSD68 with noise level of 15.}}
\begin{tabular}{cccccc}
\hline
Windows sizes & 4$\times$4   & 6$\times$6   & 8$\times$8  & 48$\times$48  & 48$\times$48 and 96$\times$96 (Ours) \\ \hline
PSNR(dB)        & 31.87 & 31.88 & 31.88 & 31.97 & 31.99       \\ \hline
\end{tabular}
\end{table*}


\subsection{Transformer direction enhancement block}
To make tradeoff between long- and short-distance modeling for image denoising, TDEBlock cut different windows in terms of different directions to improve denoising performance without increasing denoising time. That is, we use horizontal shift, vertical shift and non-shift in common Transformers to obtain three Transformer, i.e.,  Ho, Ve and Co, which can be shown in Fig.1. To extract richer detailed information, a heterogeneous architecture is designed. That is, Ho is set at the first, fourth and seventh layers. Ve is acted at the second, fifth and eighth layers. Also, Co is as the third and sixth layers. The design has three merits as follows. Firstly, TDEBlock is a heterogeneous network architecture, which can facilitate richer information. Secondly, taking into superiority of short-distance modeling account, windows are cut to sizes of $48 \times 48$ to extract local information. Thirdly, GTEBlocks and TDEBlock have heterogeneous architecture, which can facilitate richer information to promote denoising performance. Besides, different window sizes make a tradeoff between long- and short-distance modeling for image denoising.   Although a combination of GTEBlocks and TDEBlock can perform well in image denoising, it neglects effects of neighboring patches. To address this issue, we use a sparse technique into a feedforward network in each Ho, Ve and Co. That is, to enlarge more surrounding pixel information, we use dilated patches of $7\times7$ rather than that of $3 \times 3$  to achieve a sparse technique to capture more context information in image denoising, as shown in Fig.1. Besides, we reduce output dimension of the first fully-connected layer rather than obtained high-dimensional output of the first fully-connected layer in the common Transformers to remove redundant information for image denoising. \textcolor{black}{Additionally, its pseudocode can
be shown in Algorithm 3.}

\begin{algorithm}
\caption{\textcolor{black}{Ho/Ve/Co}}
\label{alg:example}
\SetKwInOut{Input}{Input}
\SetKwInOut{Output}{Output}
\Input{Input tensor $x \in {R^{B \times L \times C}}$}
\Output{Output tensor $z \in {R^{B \times L \times C}}$}
\SetKwFunction{Ho}{Ho}
\SetKwProg{Fn}{Ho}{:}{}
\Fn{\Ho{$x$}}{
    $x \gets LN(x)$
    
    $x \gets roll\_horizontal(x)$ //Ho
    
    $(x \gets roll\_ vertical(x))$ //Ve
    
    $(pass)$ //Co
    
    $y \gets self\_attention(x)$

    $y \gets roll\_reverse(y)$

    $y \gets y+x$

    $z \gets LN(y)$

    $z \gets select \_ pixel(z)$ //sparse technique

    $z \gets FCL(ReLU(FCL(z)))+y$ 
}
\end{algorithm}

\begin{algorithm}
\caption{\textcolor{black}{GTEBlock}}
\label{alg:example}
\SetKwInOut{Input}{Input}
\SetKwInOut{Output}{Output}
\Input{Input tensor $x \in {R^{B \times C \times H \times W}}$}
\Output{Output tensor $z \in {R^{B \times C \times H \times W}}$}
\SetKwFunction{GTEBlock}{GTEBlock}
\SetKwProg{Fn}{GTEBlock}{:}{}
\Fn{\GTEBlock{$x$}}{
    $split$ $x$ $in$ $n$ $windows$ $with$ $96\times96$
    
    $q,k,v \gets conv(x)$
    
    $q,k,v \gets img2seq((q, k, v)) $ 

    $q, k, v \gets LN((q, k, v))$

    $y\gets self\_attention(q, k, v)$

    $z \gets seq2img(y)$

    $z \gets z + x$

    $z \gets conv(ReLU(conv(z)))+z$
}
\end{algorithm}

\begin{algorithm}
\caption{\textcolor{black}{TDEBlock}}
\label{alg:example}
\SetKwInOut{Input}{Input}
\SetKwInOut{Output}{Output}
\Input{Input tensor $x \in {R^{B \times C \times H \times W}}$}
\Output{Output tensor $z \in {R^{B \times C \times H \times W}}$}
\SetKwFunction{TDEBlock}{TDEBlock}
\SetKwProg{Fn}{TDEBlock}{:}{}
\Fn{\TDEBlock{$x$}}{
    $x \gets img2seq(x)$
    $z \gets Ho(Ve(Co(Ho(Ve(Co(Ho(Ve(x))))))))$
    $z \gets seq2img(z)$
}
\end{algorithm}

\section{Experimental analysis and results}
\subsection{Experimental settings}

To fairly evaluate our HWformer, we use public synthetic noisy image datasets containing BSD500 with 432 natural images \cite{arbelaez2010contour}, DIV2K with 800 natural images \cite{agustsson2017ntire}, Flickr2K with 2,650 natural images \cite{timofte2017ntire} and WED with 4,744 natural images \cite{ma2016waterloo}, and real noisy image datasets containing SIDD-Medium dataset\cite{abdelhamed2018high} with 320 natural images of $5328\times3000$  resolution \cite{abdelhamed2018high} to train our HWformer, respectively. For synthetic noisy image denoising, to accelerate training denoising, we randomly crop each image into 48 image patches with sizes of $96\times 96$ and total image patches are 414,048 for each epoch. Besides, to keep diversity of synthetic noisy image training datasets, we used the same data augmentation ways as Ref. \cite{tian2024cross} to augment dataset above. For real noisy image denoising, we crop each image into 300 image patches with sizes of $192\times 192$. Other training settings are as same as synthetic image denoising. Besides, we use the following parameters to training denoising models for syntenic and real noisy image denoising. And batch size is set to 8, the number of epochs is 28. Learning rate is initialized to 1e-4 and decays to half at the 15th, 22th, 24th, 25th, 26th, 27th and 28th epochs. Adam optimizer \cite{kingma2014adam}  with  $\beta1$=0.9 and $\beta2$ =0.99. All experiments were conducted on a PC with an Ubuntu 20.04, an AMD EPYC 7502P. The PC has a 32-core CPU, 128GB of RAM and an Nvidia GeForce GTX 3090 GPU. To accelerate training speed of image denoising method, the GPU, Nvidia CUDA version 11.1 and cuDNN version 8.04 are used.
\subsection{Ablation study}

\begin{table*}
\centering
\caption{\textcolor{black}{Average PSNR(dB) of eleven gray image denoising methods on Set12\cite{zhang2017beyond} with different noise levels of 15, 25 and 50.}}
\scalebox{0.8}{

\begin{tabular}{cccccccccccccc} 
 \hline
 Images                  & C.man                                & House                                & Peppers                              & Starfish                             & Monarch                              & Airplane                                              & Parrot                                                                                                  & Barbara                              & Boat                                 & Man                                                   & Couple                               & Average                               \\ 
 \hline
    Noise level             & \multicolumn{12}{c}{$\sigma=15$}                                                                                                                                                                                                                                                                                                                                                                                                                                                                                                                                                              \\ 
 \hline
     BM3D                    & 31.91                                & 34.93                                & 32.69                                & 31.14                                & 31.85                                & 31.07                                                 & 31.37                                                                                                  & 33.10                                & 32.13                                & 31.92                                                 & 32.10                                & 32.20                                 \\
     TNRD                    & 32.19                                & 34.53                                & 33.04                                & 31.75                                & 32.56                                & 31.46                                                 & 31.63                                                                                                  & 32.13                                & 32.14                                & 32.23                                                 & 32.11                                & 32.34                                 \\
     DnCNN                   & 32.61                                & 34.97                                & 33.30                                & 32.20                                & 33.09                                & 31.70                                                 & 31.83                                                                                                  & 32.64                                & 32.42                                & 32.46                                                 & 32.47                                & 32.69                                 \\
     FFDNet                  & 32.43                                & 35.07                                & 33.25                                & 31.99                                & 32.66                                & 31.57                                                 & 31.81                                                                                                  & 32.54                                & 32.38                                & 32.41                                                 & 32.46                                & 32.60                                 \\
     FOCNet                  & 32.71                                & 35.44                                & 33.41                                & 32.40                                & 33.29                                & 31.82                                                 & 31.98                                                                                                  & 33.09                                & 32.62                                & 32.56                                                 & 32.64                                & 32.91                                 \\
     \textcolor{black}{RDDCNN}                  & 32.61                                & 35.01                                & 33.31                                & 32.13                                & 33.13                                & 31.67                                                 & 31.93                                                                                                  & 32.62                                & 32.42                                & 32.38                                                 & 32.46                                & 32.70                                 \\
     DGAL                    & 32.73                                & 35.83                                & 33.50                                & 32.57                                & 33.56                                & 31.96                                                 & 32.08                                                                                                  & 33.81                                & 32.68                                & 32.61                                                 & 32.73                                & 33.10                                 \\
     \textcolor{black}{CTNet}          &   \textcolor[rgb]{0,0.69,0.941}{32.82}                                 &   35.86                &   33.69                &   \textcolor[rgb]{0,0.69,0.941}{32.65}                &   33.53                &   \textcolor[rgb]{0,0.69,0.941}{32.07}                &   32.21                                                                  &   33.87                &   32.75                                 &   32.61                                 &   32.77                                                    &   33.20                 \\
     
     SwinIR                  & \textcolor{red}{32.93}               & \textcolor[rgb]{0,0.69,0.941}{36.00} & \textcolor[rgb]{0,0.69,0.941}{33.72} & 32.59 & \textcolor[rgb]{0,0.69,0.941}{33.66} & \textcolor{red}{32.12}                  & \textcolor[rgb]{0,0.69,0.941}{32.22}                                                  & \textcolor[rgb]{0,0.69,0.941}{33.97} & \textcolor[rgb]{0,0.69,0.941}{32.80} & \textcolor{red}{32.68}                                & \textcolor[rgb]{0,0.69,0.941}{32.84} & \textcolor[rgb]{0,0.69,0.941}{33.23}  \\
     HWformer (Ours)          & \textcolor{red}{32.93}               & \textcolor{red}{36.21}               & \textcolor{red}{33.80}               & \textcolor{red}{32.78}               & \textcolor{red}{33.68}               & 32.06                                & \textcolor{red}{32.25}                                                & \textcolor{red}{34.08}               & \textcolor{red}{32.86}               & \textcolor[rgb]{0,0.69,0.941}{32.65}                  & \textcolor{red}{32.86}               & \textcolor{red}{33.29}                \\ 
 \hline
    Noise level             & \multicolumn{12}{c}{$\sigma=25$}                                                                                                                                                                                                                                                                                                                                                                                                                                                                                                                                                              \\ 
 \hline
     BM3D                    & 29.45                                & 32.85                                & 30.16                                & 28.56                                & 29.25                                & 28.42                                                 & 28.93                                                                                                & 30.71                                & 29.90                                & 29.61                                                 & 29.71                                & 29.78                                 \\
     TNRD                    & 29.72                                & 32.53                                & 30.57                                & 29.02                                & 29.85                                & 28.88                                                 & 29.18                                                                                                 & 29.41                                & 29.91                                & 29.87                                                 & 29.71                                & 29.88                                 \\
     DnCNN                   & 30.18                                & 33.06                                & 30.87                                & 29.41                                & 30.28                                & 29.13                                                 & 29.43                                                                                               & 30.00                                & 30.21                                & 30.10                                                 & 30.12                                & 30.25                                 \\
     FFDNet                  & 30.10                                & 33.28                                & 30.93                                & 29.32                                & 30.08                                & 29.04                                                 & 29.44                                                                                               & 30.01                                & 30.25                                & 30.11                                                 & 30.20                                & 30.31                                 \\
     N\textsuperscript{3}Net & 30.08                                & 33.25                                & 30.90                                & 29.55                                & 30.45                                & 29.02                                                 & 29.45                                                                                                 & 30.22                                & 30.26                                & 30.12                                                 & 30.12                                & 30.31                                 \\
     FOCNet                  & 30.35                                & 33.63                                & 31.00                                & 29.75                                & 30.49                                & 29.26                                                 & 29.58                                                                                                  & 30.74                                & 30.46                                & 30.22                                                 & 30.40                                & 30.53                                 \\
     \textcolor{black}{RDDCNN}                  & 30.20                                & 33.13                                & 30.82                                & 29.38                                & 30.36                                & 29.05                                                 & 29.53                                                                                                 & 30.03                                & 30.19                                & 30.05                                                 & 30.10                                & 30.26                                 \\
     DGAL                    & 30.36                                & 33.88                                & 31.18                                & 29.99                                & 30.77                                & 29.40                                                 & 29.65                                                                                                  & 31.56                                & 30.51                                & 30.28                                                 & 30.45                                & 30.73                                 \\
     \textcolor{black}{CTNet}          &   30.40        &    33.86                 &    31.33                 &    \textcolor[rgb]{0,0.69,0.941}{30.03}                 &    30.68                 &    29.50                 &    29.73                                                                   &    31.62                 &    30.54                                  &    30.27                                  &    30.49                                                     &    30.77                  \\
     SwinIR                  & \textcolor{red}{30.53}               & \textcolor[rgb]{0,0.69,0.941}{33.99} & \textcolor[rgb]{0,0.69,0.941}{31.35} & 29.98 & \textcolor[rgb]{0,0.69,0.941}{30.89} & \textcolor{red}{29.55}                                & \textcolor[rgb]{0,0.69,0.941}{29.77}                                   & \textcolor[rgb]{0,0.69,0.941}{31.70} & \textcolor[rgb]{0,0.69,0.941}{30.63} & \textcolor{red}{30.33}\textcolor[rgb]{0,0.69,0.941}{} & \textcolor[rgb]{0,0.69,0.941}{30.59} & \textcolor[rgb]{0,0.69,0.941}{31.85}  \\
     HWformer (Ours)          & \textcolor[rgb]{0,0.69,0.941}{30.51} & \textcolor{red}{34.21}               & \textcolor{red}{31.41}               & \textcolor{red}{30.23}               & \textcolor{red}{30.98}               & 29.48                  & \textcolor{red}{29.83}                                                               & \textcolor{red}{31.83}               & \textcolor{red}{30.68}               & \textcolor[rgb]{0,0.69,0.941}{30.31}\textcolor{red}{} & \textcolor{red}{30.61}               & \textcolor{red}{30.92}                \\ 
 \hline
    Noise level             & \multicolumn{12}{c}{$\sigma=50$}                                                                                                                                                                                                                                                                                                                                                                                                                                                                                                                                                                         \\ 
 \hline
     BM3D                    & 26.13                                & 29.69                                & 26.68                                & 25.04                                & 25.82                                & 25.10                                                 & 25.90                                                                                        & 27.22                                & 26.78                                & 26.81                                                 & 26.46                                & 26.51                                 \\
     TNRD                    & 26.62                                & 29.48                                & 27.10                                & 25.42                                & 26.31                                & 25.59                                                 & 26.16                                                                                                  & 25.70                                & 26.94                                & 26.98                                                 & 26.50                                & 26.62                                 \\
     DnCNN                   & 27.03                                & 30.00                                & 27.32                                & 25.70                                & 26.78                                & 25.87                                                 & 26.48                                                                                                 & 26.22                                & 27.20                                & 27.24                                                 & 26.90                                & 26.98                                 \\
     FFDNet                  & 27.05                                & 30.37                                & 27.54                                & 25.75                                & 26.81                                & 25.89                                                 & 26.57                                                                                                  & 26.45                                & 27.33                                & 27.29                                                 & 27.08                                & 27.10                                 \\
     N\textsuperscript{3}Net & 27.14                                & 30.50                                & 27.58                                & 26.00                                & 27.03                                & 25.75                                                 & 26.50                                                                                                  & 27.01                                & 27.32                                & 27.33                                                 & 27.04                                & 27.2                                 \\
     FOCNet                  & 27.36                                & 30.91                                & 27.57                                & 26.19                                & 27.10                                & 26.06                                                 & 26.75                                                                                                  & 27.60                                & 27.53                                & 27.42                                                 & 27.39                                & 27.44                                 \\
     \textcolor{black}{RDDCNN}                  & 27.16                                & 30.21                                & 27.38                                & 25.72                                & 26.84                                & 25.88                                                 & 26.53                                                                                                  & 26.36                                & 27.23                                & 27.22                                                 & 26.88                                & 27.04                                 \\
     DGAL                    & 27.30                                & 31.09                                & 27.78                                & 26.37                                & 27.16                                & 26.16                                                 & 26.76                                                                                                 & 28.27                                & 27.50                                & 27.44                                                 & 27.36                                & 27.56                                 \\
     \textcolor{black}{CTNet}          &    \textcolor[rgb]{0,0.69,0.941}{27.47}                                  &    30.98                 &    \textcolor[rgb]{0,0.69,0.941}{27.92}                 &    \textcolor[rgb]{0,0.69,0.941}{26.45}                 &    27.14                 &    \textcolor{red}{26.28}                 &   26.70                &    28.29                 &   27.52        &   27.41        &   27.37                           &    27.59                  \\
     SwinIR                  & \textcolor[rgb]{0,0.69,0.941}{27.47} & \textcolor[rgb]{0,0.69,0.941}{31.25} & 27.91 & \textcolor[rgb]{0,0.69,0.941}{26.45} & \textcolor[rgb]{0,0.69,0.941}{27.24} & \textcolor[rgb]{0,0.69,0.941}{26.25}\textcolor[rgb]{0,0.69,0.941}{} & \textcolor{red}{26.95} & \textcolor[rgb]{0,0.69,0.941}{28.39} & \textcolor[rgb]{0,0.69,0.941}{27.65} & \textcolor[rgb]{0,0.69,0.941}{27.50}                  & \textcolor[rgb]{0,0.69,0.941}{27.54} & \textcolor[rgb]{0,0.69,0.941}{27.69}  \\
     HWformer (Ours)          & \textcolor{red}{27.54}               & \textcolor{red}{31.44}               & \textcolor{red}{27.96}               & \textcolor{red}{26.76}               & \textcolor{red}{27.43}               & 26.22 & \textcolor[rgb]{0,0.69,0.941}{26.92}\textcolor{red}{}  & \textcolor{red}{28.56}               & \textcolor{red}{27.72}               & \textcolor{red}{27.51}                                & \textcolor{red}{27.55}               & \textcolor{red}{27.78}                \\
 \hline
\end{tabular}
}
\end{table*}

\definecolor{Cerulean}{rgb}{0,0.69,0.941}
\begin{table}
\centering
\caption{Average PSNR(dB) of different gray image denoising methods on BSD68\cite{martin2001database} and Urban100\cite{huang2015single} with different noise levels of 15, 25 and 50.}
\scalebox{0.8}{
\begin{tblr}{
  cells = {c},
  cell{1}{2} = {c=3}{},
  cell{1}{5} = {c=3}{},
  cell{10}{2} = {fg=Cerulean},
  cell{12}{2} = {fg=Cerulean},
  cell{13}{3} = {fg=Cerulean},
  cell{13}{4} = {fg=red},
  cell{13}{5} = {fg=Cerulean},
  cell{13}{6} = {fg=Cerulean},
  cell{13}{7} = {fg=Cerulean},
  cell{14}{2} = {fg=red},
  cell{14}{3} = {fg=red},
  cell{14}{4} = {fg=Cerulean},
  cell{14}{5} = {fg=red},
  cell{14}{6} = {fg=red},
  cell{14}{7} = {fg=red},
  hline{1,15} = {-}{0.08em},
  hline{2-3} = {-}{},
}
Methods     & BSD68~      &             &             & Urban100~   &             &             \\
Noise levels & 15          & 25          & 50          & 15          & 25          & 50          \\
DnCNN       & 31.73       & 29.23       & 26.23       & 32.64       & 29.95       & 26.23       \\
FFDNet      & 31.63       & 29.19       & 26.29       & 32.43       & 29.92       & 26.52       \\
IRCNN       & 31.63       & 29.15       & 26.19       & 32.46       & 29.80       & 26.22       \\
FOCNet      & 31.83       & 29.38       & 26.50       & 33.15       & 30.64       & 27.40       \\
DRUNet      & 31.91       & 29.48       & 26.59       & 33.44       & 31.11       & 27.96       \\
DAGL        & 31.93       & 29.46       & 26.51       & 33.79       & 31.39       & 27.97       \\
\textcolor{black}{RDDCNN}      & 31.76       & 29.27       & 26.30       & -           & -           & -           \\
CSformer    & 
  31.97
   & 29.51       & 26.60       & -           & -           & -           \\
\textcolor{black}{CTNet} &31.94 &29.46 & 26.49 & 33.72 &31.28 & 27.80 \\
SwinIR      & 
  31.97
   & \textcolor[rgb]{0,0.69,0.941}{29.50}       & 26.58       & 33.70       & 31.30       & 27.98       \\
Restormer   & 31.96       & 
  29.52
   & 
  26.62
   & 
  33.79
   & 
  31.46
   & 
  28.29
   \\
HWformer (Ours)        & 
  31.99
   & 
  29.54
   & 
  26.61
   & 
  33.94
   & 
  31.61
   & 
  28.35
   
\end{tblr}
}
\end{table}

\begin{table*}
\centering
\caption{Average PSNR(dB) and LPIPS of different color image denoising methods on McMaster\cite{zhang2011color} and Urban100 \cite{huang2015single} with different noise levels of 15, 25 and 50.}
\begin{tabular}{ccccccccccc} 
\toprule
DataSets                   & $\sigma$                   & Metrics & DnCNN & DRUNet & IPT   & CTNet & SwinIR                               & Restormer                            & EDT-B                                & HWformer (Ours)                                  \\ 
\hline
\multirow{6}{*}{McMaster} & \multirow{2}{*}{15} & PSNR   & 33.45 & 35.40  & /     & 35.54 & \textcolor[rgb]{0,0.69,0.941}{35.61} & \textcolor[rgb]{0,0.69,0.941}{35.61} & \textcolor[rgb]{0,0.69,0.941}{35.61} & \textcolor{red}{35.64}                \\
                          &                     & LPIPS  & 0.147 & 0.116  & /     & 0.116 & \textcolor[rgb]{0,0.69,0.941}{0.111} & 0.112                                & 0.115                                & \textcolor{red}{0.110}                \\ 
\cline{2-11}
                          & \multirow{2}{*}{25} & PSNR   & 31.52 & 33.14  & /     & 33.21 & 33.20                                & \textcolor[rgb]{0,0.69,0.941}{33.34} & \textcolor[rgb]{0,0.69,0.941}{33.34} & \textcolor{red}{33.36}                \\
                          &                     & LPIPS  & 0.196 & 0.159  & /     & 0.159 & \textcolor{red}{0.153}               & \textcolor{red}{0.153}               & \textcolor[rgb]{0,0.69,0.941}{0.155} & \textcolor{red}{0.153}                \\ 
\cline{2-11}
                          & \multirow{2}{*}{50} & PSNR   & 28.62 & 30.08  & 29.98 & 30.02 & 30.22                                & \textcolor{red}{30.30}               & \textcolor[rgb]{0,0.69,0.941}{30.25} & 30.24                                 \\
                          &                     & LPIPS  & 0.287 & 0.229  & 0.235 & 0.234 & 0.225                                & \textcolor{red}{0.220}               & 0.225                                & \textcolor[rgb]{0,0.69,0.941}{0.223}  \\ 
\hline
\multirow{6}{*}{Urban100} & \multirow{2}{*}{15} & PSNR   & 32.98 & 34.81  & /     & 35.12 & 35.13                                & 35.13                                & \textcolor[rgb]{0,0.69,0.941}{35.22} & \textcolor{red}{35.26}                \\
                          &                     & LPIPS  & 0.112 & 0.084  & /     & 0.085 & 0.082                                & \textcolor{red}{0.080}               & 0.083                                & \textcolor[rgb]{0,0.69,0.941}{0.081}  \\ 
\cline{2-11}
                          & \multirow{2}{*}{25} & PSNR   & 30.81 & 32.60  & /     & 32.85 & 32.90                                & 32.96                                & \textcolor[rgb]{0,0.69,0.941}{33.07} & \textcolor{red}{33.10}                \\
                          &                     & LPIPS  & 0.155 & 0.117  & /     & 0.118 & 0.114                                & 0.113                                & 0.114                                & \textcolor{red}{0.112}                \\ 
\cline{2-11}
                          & \multirow{2}{*}{50} & PSNR   & 27.59 & 29.61  & 29.71 & 29.73 & 29.82                                & 30.02                                & \textcolor{red}{30.16}               & \textcolor[rgb]{0,0.69,0.941}{30.14}  \\
                          &                     & LPIPS  & 0.244 & 0.173  & 0.177 & 0.176 & 0.171                                & 0.166                                & 0.166                                & \textcolor{red}{0.163}                \\
\bottomrule
\end{tabular}
\end{table*}

\begin{table*}
\centering
\caption{\textcolor{black}{AVERAGE SSIM and FSIM OF DIFFERENT COLOR IMAGE DENOISING METHODS ON CBSD68 \cite{martin2001database}, KODAK24 \cite{franzen1999kodak} WITH DIFFERENT NOISE
LEVELS OF 15, 25 AND 50.}}
\begin{tabular}{cccccccc} 
\toprule
\multicolumn{2}{c}{Datasets}                                   & \multicolumn{3}{c}{CBSD68}                                                                                                                                                                                          & \multicolumn{3}{c}{Kodak24}                                                                                                                                                              \\ 
\hline
\multicolumn{2}{c}{Noise levels}                               & 15                                                                        & 25                                                                        & 50                                                          & 15                                                          & 25                                                          & 50                                                           \\ 
\hline
\multirow{2}{*}{DnCNN}  & SSIM                                & 0.9317                                                                    & 0.8863                                                                    & 0.7915                                                      & 0.9205                                                      & 0.8774                                                      & 0.7896                                                       \\
                        & FSIM                                & 0.7838                                                                    & 0.7371                                                                    & 0.6572                                                      & 0.7639                                                      & 0.7131                                                      & 0.6315                                                       \\ 
\hline
\multirow{2}{*}{FFDNet} & SSIM\textcolor[rgb]{0,0.69,0.941}{} & 0.9318                                                                    & 0.8860                                                                    & 0.7916                                                      & 0.9231                                                      & 0.8792                                                      & 0.7930                                                       \\
                        & FSIM\textcolor[rgb]{0,0.69,0.941}{} & 0.7818                                                                    & 0.7330                                                                    & 0.6479                                                      & 0.7636                                                      & 0.7090                                                      & 0.6215                                                       \\ 
\hline
\multirow{2}{*}{DRUNet} & SSIM\textcolor[rgb]{0,0.69,0.941}{} & \textcolor[rgb]{0,0.69,0.941}{0.9373}                                     & \textcolor[rgb]{0,0.69,0.941}{0.8963}                                     & 0.8139                                                      & 0.9304                                                      & 0.8931                                                      & 0.8205                                                       \\
                        & FSIM                                & 0.7915                                                                    & 0.7479                                                                    & 0.6752                                                      & 0.7776                                                      & 0.7302                                                      & 0.6557                                                       \\ 
\hline
\multirow{2}{*}{IPT}    & SSIM                                & /                                                                         & /                                                                         & 0.8090                                                      & /                                                           & /                                                           & 0.8155                                                       \\
                        & FSIM                                & /                                                                         & /                                                                         & 0.6689                                                      & /                                                           & /                                                           & 0.6507                                                       \\ 
\hline
\multirow{2}{*}{CTNet}  & SSIM                                & 0.9378                                                                    & 0.8963                                                                    & 0.8107                                                      & 0.9309                                                      & 0.8930                                                      & 0.8168                                                       \\
                        & FSIM                                & 0.7929                                                                    & 0.7487                                                                    & 0.6735                                                      & 0.7790                                                      & 0.7315                                                      & 0.6547                                                       \\ 
\hline
\multirow{2}{*}{SwinIR} & SSIM\textcolor{red}{}               & \textcolor{red}{0.9384}                                                   & \textcolor{red}{0.8977}                                                   & \textcolor[rgb]{0,0.69,0.941}{0.8154}                       & 0.9316                                                      & \textcolor[rgb]{0,0.69,0.941}{0.8945}                       & 0.8220                                                       \\
                        & FSIM\textcolor{red}{}               & \textcolor{red}{0.7940}                                                   & \textcolor{red}{0.7505}                                                   & \textcolor[rgb]{0,0.69,0.941}{0.6779}                       & \textcolor[rgb]{0,0.69,0.941}{0.7805}                       & \textcolor[rgb]{0,0.69,0.941}{0.7338}                       & \textcolor[rgb]{0,0.69,0.941}{0.6600}                        \\ 
\hline
\multirow{2}{*}{EDT-B}  & SSIM                                & 0.9352                                                                    & 0.8933                                                                    & 0.8112                                                      & \textcolor[rgb]{0,0.69,0.941}{0.9317}                       & 0.8931                                                      & \textcolor{red}{0.8230}                                      \\
                        & FSIM                                & 0.7929                                                                    & 0.7490                                                                    & 0.6760                                                      & 0.7794                                                      & 0.7323                                                      & 0.6589                                                       \\ 
\hline
\multirow{2}{*}{HWformer (Ours)}   & SSIM                                & \textcolor{red}{0.9384}                                                   & \textcolor{red}{0.8977}                                                   & \textcolor{red}{0.8157}                                     & \textcolor{red}{0.9320}\textcolor[rgb]{0,0.69,0.941}{}      & \textcolor{red}{0.8952}                                     & \textcolor[rgb]{0,0.69,0.941}{0.8222}                        \\
                        & FSIM                                & \textcolor[rgb]{0,0.69,0.941}{0.7936}\textcolor[rgb]{0.051,0.051,0.051}{} & \textcolor[rgb]{0,0.69,0.941}{0.7502}\textcolor[rgb]{0.051,0.051,0.051}{} & \textcolor{red}{0.6784}\textcolor[rgb]{0.051,0.051,0.051}{} & \textcolor{red}{0.7810}\textcolor[rgb]{0.051,0.051,0.051}{} & \textcolor{red}{0.7347}\textcolor[rgb]{0.051,0.051,0.051}{} & \textcolor{red}{0.6613}\textcolor[rgb]{0.051,0.051,0.051}{}  \\
\bottomrule
\end{tabular}
\end{table*}

\begin{table*}
\centering
\caption{\textcolor{black}{AVERAGE PSBR OF DIFFERENT COLOR IMAGE DENOISING METHODS ON Urban100 \cite{huang2015single} WITH DIFFERENT NOISE
LEVELS OF 15, 25 AND 50.}}
\begin{tabular}{cccccccc} 
\toprule
Methods & DnCNN & DRUNet & IPT   & CTNet & SwinIR & Restormer                            & HWformer (Ours)                           \\ 
\hline
$\sigma=$15      & 34.17 & 35.38  & /     & 35.59 & 35.15  & \textcolor[rgb]{0,0.69,0.941}{35.64} & \textcolor{red}{35.72}\textcolor{red}{}  \\
$\sigma=$25      & 32.89 & 34.01  & /     & 34.12 & 34.21  & \textcolor[rgb]{0,0.69,0.941}{34.27} & \textcolor{red}{34.31}\textcolor{red}{}  \\
$\sigma=$50      & 31.43 & 32.58  & 32.55 & 32.52 & 32.70  & \textcolor{red}{32.85}               & \textcolor[rgb]{0,0.69,0.941}{32.79}     \\
\bottomrule
\end{tabular}
\end{table*}

\begin{table*}
\centering
\caption{\textcolor{black}{AVERAGE color difference OF DIFFERENT COLOR IMAGE DENOISING METHODS ON McMaster \cite{zhang2011color} WITH DIFFERENT NOISE
LEVELS OF 15, 25 AND 50.}}
\begin{tabular}{cccccccc} 
\toprule
Methods & DnCNN  & IPT    & CTNet  & SwinIR                                & Restormer & EDT-B  & HWformer (Ours)           \\ 
\hline
$\sigma=15$      & 0.1233 & /      & 0.0483 & \textcolor[rgb]{0,0.69,0.941}{0.0409} & 0.0545    & 0.0483 & \textcolor{red}{0.0295}  \\
 $\sigma=25$     & 0.1586 & /      & 0.0641 & \textcolor[rgb]{0,0.69,0.941}{0.0603} & 0.0852    & 0.0722 & \textcolor{red}{0.0430}  \\
$\sigma=50$      & 0.2547 & 0.2137 & 0.1519 & \textcolor[rgb]{0,0.69,0.941}{0.1148} & 0.1452    & 0.1472 & \textcolor{red}{0.0855}  \\
\bottomrule
\end{tabular}
\end{table*}

\begin{table}[]
\caption{\textcolor{black}{Average PSNR(DB) of different color image denoising methods on SPARCS \cite{hughes2014automated} with different noise levels of 15, 25 and 50.}}
\centering
\begin{tabular}{cccc}
\hline
Methods & 15    & 25    & 50    \\ \hline
DnCNN   & 35.98 & 34.19 & 31.69 \\
DRUNet  & 37.22 & 35.16 & 32.58 \\
IPT &-- &-- & 32.39 \\
CTNet & 37.25 & 35.13 &32.42 \\
SwinIR  & \textcolor{Cerulean}{37.32} & \textcolor{Cerulean}{35.23} & \textcolor{red}{32.63} \\
HWformer (Ours)    & \textcolor{red}{37.35} & \textcolor{red}{35.25} & \textcolor{Cerulean}{32.61} \\ \hline
\end{tabular}
\end{table}

\begin{table*}
\centering
\caption{\textcolor{black}{PSNR (dB) results of different methods on CC.}}
\begin{tabular}{ccccccc} 
\toprule

 Settings                                & CBM3D & TID & DnCNN & RDDCNN & DeCapsGAN & HWformer (Ours)
\\ 
\hline
\multirow{3}{*}{Canon 5D ISO = 3200}    & \textcolor{red}{39.76}               & 37.22 & \textcolor[rgb]{0,0.69,0.941}{37.26}                                & 37.00 & 35.74                                & 36.60                   \\
                                       & 36.40                                & 34.54 & 34.13    & 33.88                                & \textcolor[rgb]{0,0.69,0.941}{37.02} & \textcolor{red}{37.22}  \\
                                       & 36.37 & 34.25 & 34.09    & 33.82                                & \textcolor[rgb]{0,0.69,0.941}{36.74}                                & \textcolor{red}{38.93}  \\ 
\hline
\multirow{3}{*}{Nikon D600 ISO = 3200} & 34.18 & 32.99 & 33.62                                & 33.24                                & \textcolor{red}{35.71}                                & \textcolor[rgb]{0,0.69,0.941}{35.29}  \\
                                       & 35.07 & 34.20 & 34.48    & 33.76                                & \textcolor[rgb]{0,0.69,0.941}{35.83}                                & \textcolor{red}{35.93}  \\
                                       & \textcolor[rgb]{0,0.69,0.941}{37.13}                                & 35.58 & 35.41  & 34.91                                & 36.93                                & \textcolor{red}{40.51}  \\ 
\hline
\multirow{3}{*}{Nikon D800 ISO = 1600} & 36.81                                & 34.49 & 37.95                                & 35.47                                & \textcolor[rgb]{0,0.69,0.941}{38.41} & \textcolor{red}{38.97}  \\
                                       & 37.76                                & 35.19 & 36.08                                 & 34.81                                & \textcolor[rgb]{0,0.69,0.941}{39.14} & \textcolor{red}{39.46}  \\
                                       & \textcolor[rgb]{0,0.69,0.941}{37.51} & 35.26 & 35.48    & 35.71                                & 37.19                                & \textcolor{red}{38.31}  \\ 
\hline
\multirow{3}{*}{Nikon D800 ISO = 3200} & 35.05                                & 33.70 & 34.08 & \textcolor[rgb]{0,0.69,0.941}{37.20}                                & 36.93                                & \textcolor{red}{37.68}  \\
                                       & 34.07                                & 31.04 & 33.70    & 32.89                                & \textcolor[rgb]{0,0.69,0.941}{36.85} & \textcolor{red}{37.37}  \\
                                       & 34.42                                & 33.07 & 33.31  & 32.91                                & \textcolor[rgb]{0,0.69,0.941}{36.85}                               & \textcolor{red}{39.21}  \\ 
\hline
\multirow{3}{*}{Nikon D800 ISO = 6400} & 31.13                                & 29.40 & 29.83  & 29.86                                & \textcolor{red}{33.32}                                & \textcolor[rgb]{0,0.69,0.941}{33.13}  \\
                                       & 31.22                                & 29.86 & 30.55 & 29.97                                & \textcolor[rgb]{0,0.69,0.941}{31.81}   & \textcolor{red}{33.12}  \\
                                       & 30.97                                & 29.21 & 30.09  & 29.63                                & \textcolor{red}{33.67}    & \textcolor[rgb]{0,0.69,0.941}{33.32}  \\ 
\hline
Average                                & 35.19                                & 33.63 & 33.86                                & 33.41                                & \textcolor[rgb]{0,0.69,0.941}{36.73} & \textcolor{red}{36.89}  \\
\bottomrule
\end{tabular}
\end{table*}

\begin{table*}
  \caption{Average PSNR(dB) of different real image denoising methods on SIDD\cite{abdelhamed2018high}.}
  \label{tab:freq}
\scalebox{0.90}{
  \begin{tabular}{ccccccccccc}
    \toprule
Methods & DnCNN & RIDNet& AINDNet & SADNet & DANet & DeamNet & CycleISP & DAGL & MPRNet & HWformer (Ours) \\
    \midrule
PSNR(dB)   & 23.66    & 38.71      & 38.95      & 39.46      & 39.47     & 39.35       & 39.52        & 38.94    & \textcolor{cyan}{39.71}      & \textcolor{red}{39.72}                   \\
  \bottomrule
\end{tabular}}
\end{table*}

\begin{table*}
\centering
  \caption{\textcolor{black}{Parameters, FLOPs and running time of different methods for image denoising with different sizes.}}
  \label{tab:freq}
  \begin{tabular}{cccccc}
    \toprule
Methods      & IPT  & SwinIR & Restormer & EDT-B  & HWformer (Ours)  \\
    \midrule
Parameters       & 115.31M & 11.50M    & 26.11M        & 11.33M     & 40.06M           \\\hline
Image sizes   & \multicolumn{5}{c}{96$\times$96}\\\hline
FLOPs        & 231.07G & 105.04G   & 19.83G        & 104.02G    & 42.70G           \\
Running time  & 0.25s   & 0.060s    & 0.046s        & 0.132s     & 0.014s           \\\hline
Image sizes   & \multicolumn{5}{c}{192$\times$192} \\\hline
FLOPs        & 528.16G & 420.12G   & 79.30G        & 416.04G    & 170.42           \\
Running time  & 1.055s  & 0.589s    & 0.064s        & 1.09s      & 0.051s\\
  \bottomrule
\end{tabular}
\end{table*}

It is known that existing short-distance modeling will lose global information to reduce effect of image processing. To overcome the drawback, we design a heterogeneous architecture containing two GTEBlocks and one TDEBlock with proposed sparse technique to build bridge for interacting between short- and long-distance modeling in image denoising. Thus, in this subsection, we give principle and effectiveness of key techniques, i.e., two GTEBlocks, TDEBlock and sparse technique in the HWformer as follows. 

\begin{figure*}[h]
  \centering
  \includegraphics[width=\linewidth]{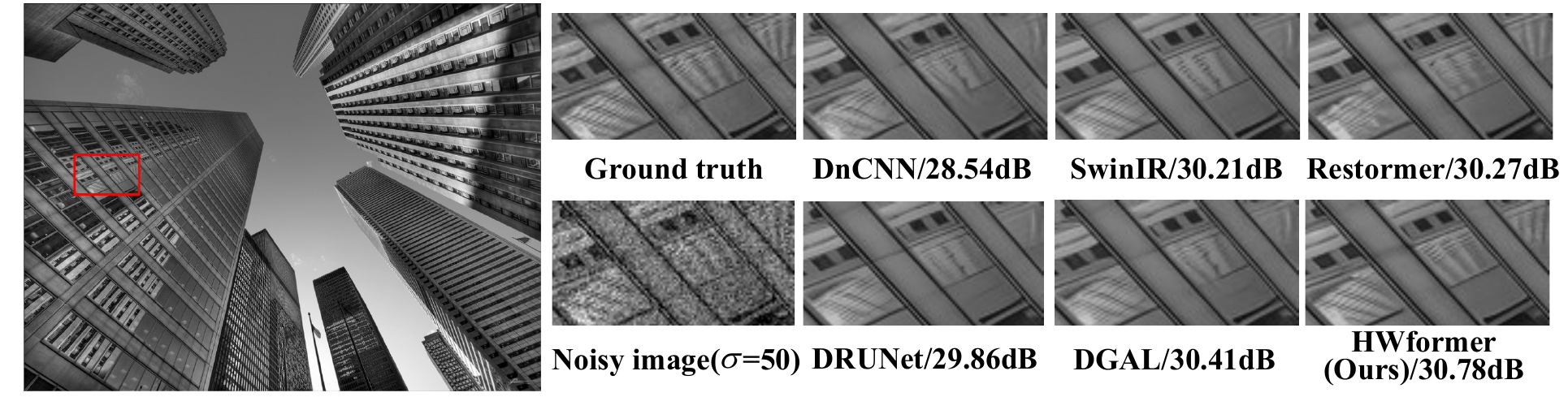}
  \caption{Visual comparisons with state-of-the-art methods on \textcolor{black}{gray} image denoising. The sample comes from Urban100\cite{huang2015single}.}
\end{figure*}

\begin{figure*}[h]
  \centering
  \includegraphics[width=\linewidth]{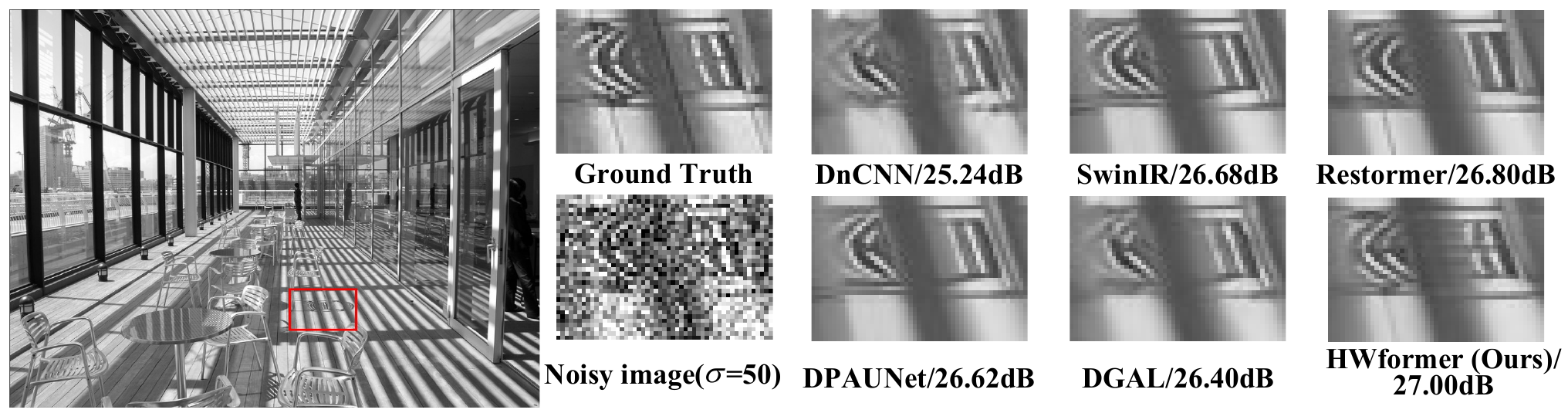}
  \caption{Visual comparisons with state-of-the-art methods on \textcolor{black}{gray} image denoising. The samples come from Urban100\cite{huang2015single}.}
\end{figure*}


\begin{figure}[h]
  \centering
  \includegraphics[width=\linewidth]{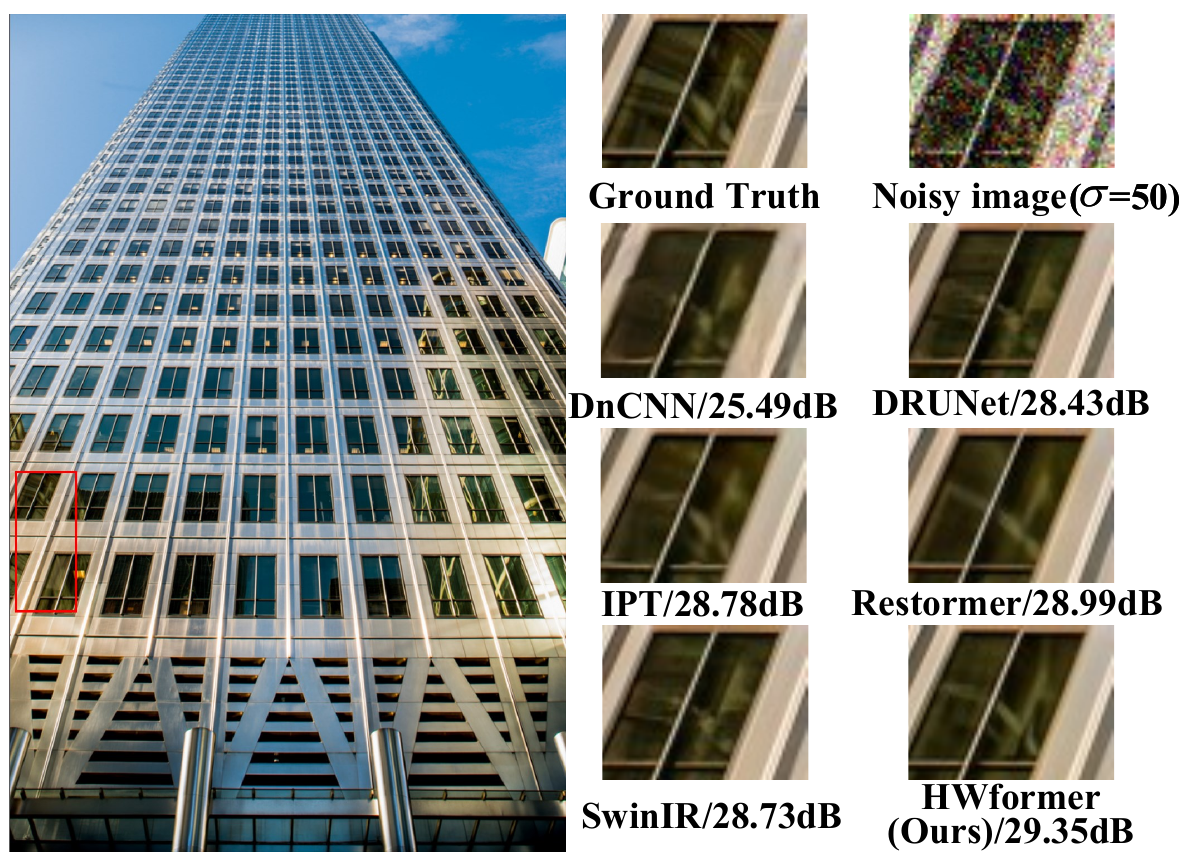}
  \caption{XVisual comparisons with state-of-the-art methods on color image denoising. The samples come from Urban100\cite{huang2015single}.}
\end{figure}

\begin{figure}[h]
  \centering
  \includegraphics[width=\linewidth]{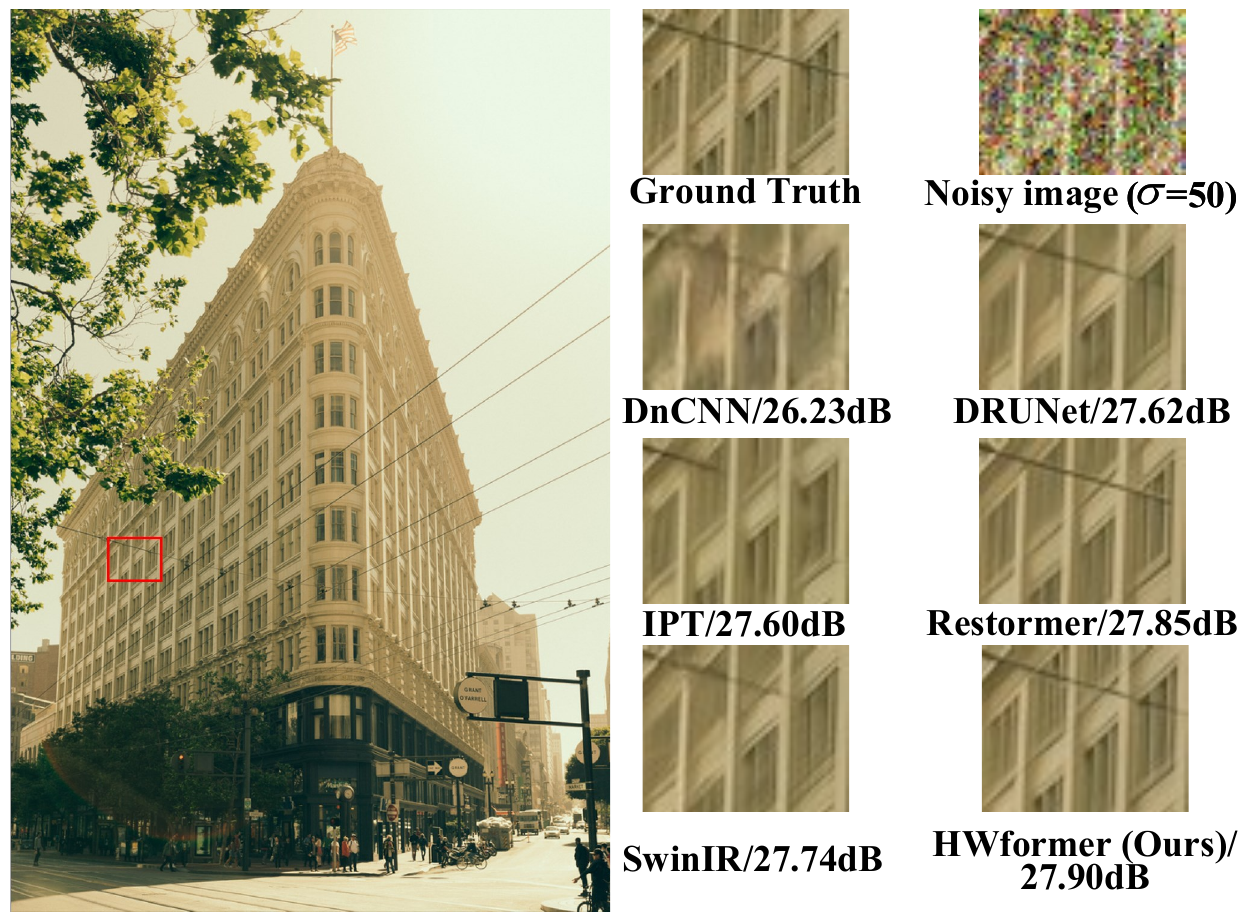}
  \caption{Visual comparisons with state-of-the-art methods on color image denoising. The samples come from Urban100\cite{huang2015single}.}
\end{figure}

\begin{figure*}[h]
  \centering
  \includegraphics[width=\linewidth]{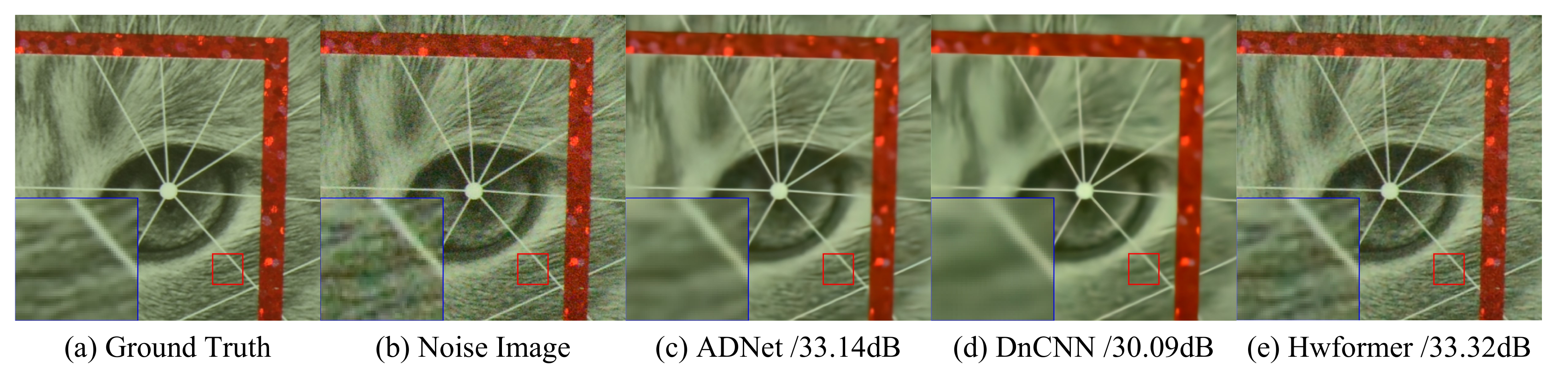}
  \caption{Visual comparisons with state-of-the-art methods on real image denoising. The samples come from CC\cite{nam2016holistic}.}
\end{figure*}

\textcolor{black}{Global-window Transformer enhancement block: Taking into inferiority of short-distance modeling account, we use windows of $96 \times 96$ rather than that of $48\times 48$ as an input of GTEBlock to capture more global information. ID 7 of GTEBlock has obtained an improvement of 0.08dB in terms of PSNR than that ID 1 for image denoising in TABLE I, which shows superiority of enlarging windows in our HWformer for image denoising. Also, improvement of 0.1dB is not easy for image denoising. Thus, it has a reference value for improving Transformer for image denoising. Because Transformer mechanism can only depend on relations of pixels to extract salient information \cite{chen2021pre},  three fully connected layers in the Transformer mechanism depend on sizes of given patches, which will result in high complexities when given patches is big. To overcome this issue, we use three convolutional layers with sizes of fixed kernels rather than three fully connected layers to eliminate native effects of three fully connected layers in terms of the number of parameters for image denoising. Its effectiveness can be verified as follows. We assume that parameters of each fully-connected layer are $C\times p^2 \times C \times p^2$, where $C$ is channel number, $p$ is size of given patches ($p=6$ in the GTEBlock). According to that, we see that parameters of each fully-connected layer are strong relation with $p$. However, parameters of each convolutional layer are $C \times C \times k \times k$ , where $k$ denotes kernel size ($k=3$). According to parameter computation method, we can see that the number of parameters is not affected by given patch sizes. Also, the number of parameters from each convolutional layer only takes 1/144 that of each fully connected layer, which can reduce computational costs. Besides, its performance does not have any varying via ID 5 of GTEBlock with fully connected layers (FCL) and ID 6 of GTEBlock as shown in TABLE I. We can see that GTEBlock with three fully connected layers has obtained same PSNR as GTEBlock with three convolutional layers on Urban100 \cite{huang2015single} for image denoising with noise level of 15.} 

Transformer direction enhancement block: Taking into superiority of short-distance modeling account, we design a heterogeneous architecture to build a bridge between long and short-distance modeling, according to relations of local areas in an image. That is, we choose windows with sizes of $48 \times 48$ as inputs of TDEBlock to capture global information, which is complementary with GTEBlock with windows of $96 \times 96$ to extract more global information, where more detailed information of winwow sizes can be given latter. ID 6 of a combination method has higher PSNR value than that of ID 4 in TABLE I. To prevent loss of some local information, we firstly design different directional shifts, i.e., Ho, Ve and Co to obtain different windows to make up eight Transformers of three kinds in the TDEBlock to facilitate richer information. That is, Ho is set to the first, fourth and seventh layers, which can be used to extract horizontal information. ID 2 has obtained higher PSNR than that of ID 1 in TABLE I, which shows effectiveness of horizontal shifts to conduct windows for image denoising. Ve is set to the second, fifth and eighth layers, which can be used to vertically extract horizontal information. PSNR of ID 3 is higher than ID 1 as listed in TABLE I, which illustrate effectiveness of vertical shifts for image denoising. Co is set to the third and sixed layers to keep original information of common Transformers in terms of enhancing relations of pixels. ID 4 is higher than ID1, ID2 and ID3 in terms of PSNR for image denoising, which shows superiority of a combination of three shifts for image denoising. Although obtained local information from Ho, Ve and Co in their internal and global information from Ho, Ve and Co in their external is complementary, this mechanism does not still inherit merit of short-distance modeling. To address this issue, we propose a sparse technique in the TDEBlock. 

\textcolor{black}{
Window size: It is known that window-based Transformer with a small window interacts less pixel content \cite{liang2021swinir}.
Thus, window sizes can limit capacity of obtained information. If window sizes, i.e.,  3$\times$3, 5$\times$5, and 7$\times$7 will cause loss information of global information. Besides, GTEBlock window requires halving operations in Section II. D, which abandons window size with odd. Thus, 4$\times$4, 6$\times$6, and 8$\times$8 can be used to conduct experiments. They have obtained poor performance than that of 96$\times$96 with big window in TABLE II. To break mentioned limitation, we choose window size of  48$\times$48 to make a designed Transformers pursue better denoising performance, according to IPT \cite{chen2021pre}. Inspired by superiority of heterogeneous networks \cite{tian2022heterogeneous}, we choose a combination of 48$\times$48 and 96$\times$96 as window sizes to make a trade-off between capturing long and short-distance information for image denoising in this paper. As illusrated in TABLE II, we can see that the proposed HWformer with windows of 48$\times$48 and 96$\times$96 has obtained higher PSNR than that of only using windows of 48$\times$48. That shows effectiveness of choosing window sizes in this paper. }

Sparse technique: It is known that a feedforward network in a Transformer only uses a fixed mapping of  $3 \times 3$ to enhance features of channels. However, it ignores effects of surrounding pixels of the fixed mapping, which will lose some local detailed information. It is known that dilated convolutions  \cite{yu2015multi} can enlarge receptive field to enlarge mapping range for capturing more context information. Inspired by that, dilated convolutional idea is first introduced into a feedforward network to achieve a sparse mechanism for obtaining more local information. That is, we use patches of $7 \times 7$ rather than that of $3 \times 3$  in dilated convolutional mapping way to achieve the sparse technique to capture more context information in image denoising, according to surrounding information of given mapping. Besides, we reduce output dimension of the first fully-connected layer (FCL) rather than obtained high-dimensional output of the first fully-connected layer in the common Transformers to remove redundant information for image denoising. ID 9 has obtained higher PSNR value than that of ID 8 in TABLE I, which shows effectiveness of sparse mechanism for extracting local information for image denoising.

According to mentioned illustrations, global-window Transformer enhancement block and Transformer direction enhancement block make up a heterogeneous architecture to facilitate more global information. Also, Transformer direction enhancement block designs different directional windows to simultaneously extract global and local information. To extract more local information, sparse technique embedded into Transformer direction enhancement block to mine more context information, according to relation of neighboring patches. Besides, the proposed dimension reduction method can ensure stability of parameters in our HWformer. Thus, the proposed method can build a bridge between long- and short-distance modeling to achieve an efficient Transformer denoiser. 

\subsection{Experimental results}
We use quantitative and qualitative analysis to test denoising performance of our HWformer. Quantitative analysis includes synthetic and real noisy image denoising, comparisons of complexity and evaluation of denoising time. Qualitative analysis is used to evaluate visual effects of predicted denoising images. 
For synthetic noisy image denoising, we use Block-matching and 3D filtering (BM3D) \cite{dabov2007image}, trainable nonlinear reaction diffusion (TNRD) \cite{chen2016trainable}, targeted image denoising (TID) \cite{luo2015adaptive}, a denoising CNN (DnCNN) \cite{zhang2017beyond}, image restoration CNN (IRCNN) \cite{zhang2017learning}, fast and flexible denoising network (FFDNet) \cite{zhang2018ffdnet},
neural nearest neighbors networks (N$^3$Net) \cite{plotz2018neural}, fractional optimal control network (FOCNet) \cite{jia2019focnet}, DRUNet \cite{zhang2021plug}, DAGL \cite{mou2021dynamic}, \textcolor{black}{robust deformed denoising CNN (RDDCNN) \cite{zhang2023robust},  cross Transformer denoising
CNN (CTNet) \cite{tian2024cross}}, CSformer \cite{yin2022csformer}, image restoration using swin Transformer (SwinIR) \cite{liang2021swinir}, Restoration Transformer (Restormer) \cite{zamir2022restormer}, image processing Transformer (IPT) \cite{chen2021pre} and encoder-decoder-based Transformer (EDT) \cite{li2021efficient} on Urban100 \cite{huang2015single}, BSD68 \cite{martin2001database} and Set12  \cite{zhang2017beyond} for gray synthetic noisy image denoising, on \textcolor{black}{CBSD68 \cite{martin2001database}, Kodak24 \cite{franzen1999kodak}, SPARCS \cite{hughes2014automated}}, McMaster \cite{zhang2011color}  and Urban100 \cite{huang2015single} for color synthetic noisy image denoising. To comprehensively validate the proposed method in image denoising, peak signal-to-noise ratio (PSNR)\cite{hore2010image}, structural similarity index (SSIM)\cite{hore2010image}, feature similarity index measure (FSIM)\cite{zhang2011fsim}, learned perceptual image patch similarity (LPIPS) \cite{zhang2018unreasonable}, \textcolor{black}{peak signal-to-blur ratio (PSBR)\cite{russo2015new} and color difference \cite{sharma2005ciede2000}} are used as metric to conduct experiments. \textcolor{black}{For gray image denoising, experiments of different methods for images denoising with single class and multiple classes are conducted. As shown in TABLE III, our HWformer method almost has obtained the best result for single image denoising with eleven classes when noise level is 15, 25 and 50, respectively. For instance, our HWformer exceeds 0.31dB than that of the second SwinIR for a Starfish image denoising with noise level of 50 in TABLE III. As shown in TABLE IV, we can see that our HWformer has nearly obtained the highest PSNR on BSD68 and Urban 100 for image denoising with multiple classes when noise level is 15, 25 and 50, respectively. For instance, our HWformer has an improvement of 0.15dB than that of the second Restormer on Urban 100 for noise level of 25 in TABLE IV.} According to mentioned illustrations, it is known that our HWformer is effective for gray image denoising. 

\textcolor{black}{For color image denoising, different methods on public datasets, i.e., McMaster, Urban100, CBSD68, Kodak24, SPARCS for image denoising are evaluated via important metric, i.e., PSNR, LPIPS, SSIM, FSIM, PSBR and color difference in TABLEs V-IX, As shown in TABLE V, we can see that our method has an improvement of 0.14dB than that of Restormer and 0.20dB than that of SwinIR on Urban100 for color synthetic noisy image denoising with noise level noise of 25. In TABLE V, we can see that our HWformer has obtained lower LPIPS than that of the second EDT-B on McMaster for image denoising with noise level of 15. In terms of a quality assessment perspective containing SSIM and FSIM, our HWformer is superior to other popular methods, i.e., DnCNN, FFDNet, DRUNet, CTNet and EDT-B. As shown in TABLE VI, we can see that our HWformer has obtained an improvement of 0.0014 than that of the second DRUNet on CBSD68 for noise level of 25 and 0.005 than that of the second SwinIR on CBSD68 for noise level of 50. In terms of detail preservation and color retention, PSBR and color difference are used to conducted experiments. As shown in TABLE VII, we can see that our HWformer has obtained the best result on Urban100 for noise levels of 15 and 25. Also, it has obtained the second result on Urban 100 for noise level of 50. For instance, our method achieves an improvement of 0.57 in PSBR compared to SwinIR on the Urban100 with the noise level of 15. As shown in TABLE VIII, our method achieves the best performance on the McMaster with noise levels of 15, 25 and 50. To test robustness of our HWformer for other applications, i.e., remote sensing, we set experiments on SPACS. In TABLE IX, we can see that the proposed HWformer has obtained the highest result for noise levels of 15 and 25. Also, it has obtained similar effects with other methods, i.e., SwinIR for noise level of 50. Thus, the proposed HWformer is effective for remote sense. According to mentioned analysis, we can see that the proposed HWformer is useful for color image denoising.}  

For real noisy image denoising, we compare with popular methods, i.e., BM3D \cite{dabov2007image}, 
TID \cite{luo2015adaptive}, DnCNN \cite{zhang2017beyond}, RIDNet \cite{anwar2019real}, VDN \cite{yue2019variational}, AINDNet \cite{kim2020transfer}, SADNet  \cite{chang2020spatial}, RDDCNN \cite{zhang2023robust}, DANet \cite{yue2020dual}, DeamNet \cite{ren2021adaptive}, CycleISP \cite{zamir2020cycleisp}, DAGL \cite{mou2021dynamic} and MPRNet \cite{multiZamir} on  \textcolor{black}{ CC \cite{nam2016holistic}} and SIDD \cite{abdelhamed2018high}. Specifically, we use given reference images (Ground truth) from chosen real noisy image datasets and obtained clean images from different noisy models to compute PSNR values \cite {hore2010image}   to test denoising performance of our HWformer. As shown in TABLE X, our method has achieved the best performance besides one setting of Nikon D800 ISO = 1600 on CC for real noisy image denoising. For instance, our method has an improvement of 3.38dB than that of the second CBM3D in terms of PSNR values of the third image with the setting of Nikon D600 ISO = 3200 for real noisy image denoising in TABLE X. As TABLE XI, we can see that the proposed HWformer is competitive in contrast with other popular methods, i.e., ADGL and MPRNet on SIDD for real noisy image denoising.Also, red and blue lines denote the best and second results in TABLEs III-XI.

To test applicability of our HWformer on real applications, we test its complexity and denoising time. To keep consistency with popular denoisng methods,i.e., Refs \cite{ji2021u2, zhao2023comprehensive}, RTX 3090 GPU is used to conduct experiments to test denoising time. As shown in TABLE XII, although our HWformer is not competitive with  Restormer and SwinIR in complexities, i.e., parameters and flops, it only takes 30\% of Restormer and 23\% of SwinIR on a noisy image with size of $96\times 96$ in terms of denoising time. Thus, it is suitable to real applications, i.e., phones and cameras. Mentioned analysis shows that our HWformer is effective for image denoising in terms of quality evaluation.  

Qualitative analysis: We choose an area of denoising images from different methods to amplify it as observation area. If observation area is clearer, its corresponding method is more effective for image denoising. Also, we choose DnCNN, SwinIR, Restormer, DRUNet, DGAL and our HWformer on BSD68 and Urban100 for noise levels of 50 to obtain gray clean images. As shown in Figs. 2 and 3, we can see that our HWformer has obtained clearer areas than that of other comparative methods. For comparisons of color visual images, we choose DnCNN, SwinIR, Restormer, DRUNet, EDT-B and our HWformer on McMaster and Urban100 for noise levels of 50 to obtain color clean images. As shown in Figs. 4 and 5, we can see that our HWformer can obtain clearer detailed information than that of other methods. For real image denoising, we can observe our HWformer is capable of restoring more details and texture information as shown in Fig. 6. Thus, our HWformer is effective for qualitative analysis. 

According to mentioned illustrations, we can see that our method has obtained comparative results for image denoising. Also, it has faster denoising time and good visual results. Thus, it is suitable to deploy on real applications, i.e., phones and cameras. 

\section{Conclusion}
In this paper, we propose a heterogeneous window Transformer (HWformer) for image denoising. HWformer first designs heterogeneous global windows to facilitate richer global information to overcome limitation of short-distance modeling. Taking into superiority of short-distance modeling account, global windows are shifted in terms of different directions to facilitate diversified information without increasing denoising time. To prevent native effects of independent patches, sparse idea is first embedded into a feed-forward network to extract more local information of neighboring patches. Our HWformer has faster denoising time, which is suitable to smart phones and cameras. In the future, we will reduce the HWformer to reduce computational costs for image denoising. 





\ifCLASSOPTIONcaptionsoff
  \newpage
\fi




%

\bibliographystyle{IEEEtran}
\bibliography{IMSC_AGL}

\begin{thebibliography}{10}
\providecommand{\url}[1]{#1}
\csname url@samestyle\endcsname
\providecommand{\newblock}{\relax}
\providecommand{\bibinfo}[2]{#2}
\providecommand{\BIBentrySTDinterwordspacing}{\spaceskip=0pt\relax}
\providecommand{\BIBentryALTinterwordstretchfactor}{4}
\providecommand{\BIBentryALTinterwordspacing}{\spaceskip=\fontdimen2\font plus
\BIBentryALTinterwordstretchfactor\fontdimen3\font minus \fontdimen4\font\relax}
\providecommand{\BIBforeignlanguage}[2]{{%
\expandafter\ifx\csname l@#1\endcsname\relax
\typeout{** WARNING: IEEEtran.bst: No hyphenation pattern has been}%
\typeout{** loaded for the language `#1'. Using the pattern for}%
\typeout{** the default language instead.}%
\else
\language=\csname l@#1\endcsname
\fi
#2}}
\providecommand{\BIBdecl}{\relax}
\BIBdecl

\bibitem{levin2011natural}
A.~Levin and B.~Nadler, ``Natural image denoising: Optimality and inherent bounds,'' in \emph{CVPR 2011}.\hskip 1em plus 0.5em minus 0.4em\relax IEEE, 2011, pp. 2833--2840.

\bibitem{li2012group}
S.~Li, H.~Yin, and L.~Fang, ``Group-sparse representation with dictionary learning for medical image denoising and fusion,'' \emph{IEEE Transactions on Biomedical Engineering}, vol.~59, no.~12, pp. 3450--3459, 2012.

\bibitem{beck2009fast}
A.~Beck and M.~Teboulle, ``Fast gradient-based algorithms for constrained total variation image denoising and deblurring problems,'' \emph{IEEE Transactions on Image Processing}, vol.~18, no.~11, pp. 2419--2434, 2009.

\bibitem{maggioni2012nonlocal}
M.~Maggioni, V.~Katkovnik, K.~Egiazarian, and A.~Foi, ``Nonlocal transform-domain filter for volumetric data denoising and reconstruction,'' \emph{IEEE Transactions on Image Processing}, vol.~22, no.~1, pp. 119--133, 2012.

\bibitem{muhammad2018efficient}
K.~Muhammad, J.~Ahmad, Z.~Lv, P.~Bellavista, P.~Yang, and S.~W. Baik, ``Efficient deep cnn-based fire detection and localization in video surveillance applications,'' \emph{IEEE Transactions on Systems, Man, and Cybernetics: Systems}, vol.~49, no.~7, pp. 1419--1434, 2018.

\bibitem{yu2020two}
Y.~Yu, Z.~Cao, Z.~Liu, W.~Geng, J.~Yu, and W.~Zhang, ``A two-stream cnn with simultaneous detection and segmentation for robotic grasping,'' \emph{IEEE Transactions on Systems, Man, and Cybernetics: Systems}, vol.~52, no.~2, pp. 1167--1181, 2020.

\bibitem{zhang2017beyond}
K.~Zhang, W.~Zuo, Y.~Chen, D.~Meng, and L.~Zhang, ``Beyond a gaussian denoiser: Residual learning of deep cnn for image denoising,'' \emph{IEEE Transactions on Image Processing}, vol.~26, no.~7, pp. 3142--3155, 2017.

\bibitem{zhang2020residual}
Y.~Zhang, Y.~Tian, Y.~Kong, B.~Zhong, and Y.~Fu, ``Residual dense network for image restoration,'' \emph{IEEE Transactions on Pattern Analysis and Machine Intelligence}, vol.~43, no.~7, pp. 2480--2495, 2020.

\bibitem{tian2021asymmetric}
C.~Tian, Y.~Xu, W.~Zuo, C.-W. Lin, and D.~Zhang, ``Asymmetric cnn for image superresolution,'' \emph{IEEE Transactions on Systems, Man, and Cybernetics: Systems}, vol.~52, no.~6, pp. 3718--3730, 2021.

\bibitem{chen2021pre}
H.~Chen, Y.~Wang, T.~Guo, C.~Xu, Y.~Deng, Z.~Liu, S.~Ma, C.~Xu, C.~Xu, and W.~Gao, ``Pre-trained image processing transformer,'' in \emph{Proceedings of the IEEE/CVF Conference on Computer Vision and Pattern Recognition}, 2021, pp. 12\,299--12\,310.

\bibitem{liang2021swinir}
J.~Liang, J.~Cao, G.~Sun, K.~Zhang, L.~Van~Gool, and R.~Timofte, ``Swinir: Image restoration using swin transformer,'' in \emph{Proceedings of the IEEE/CVF International Conference on Computer Vision}, 2021, pp. 1833--1844.

\bibitem{kim2019grdn}
D.-W. Kim, J.~Ryun~Chung, and S.-W. Jung, ``Grdn: Grouped residual dense network for real image denoising and gan-based real-world noise modeling,'' in \emph{Proceedings of the IEEE/CVF Conference on Computer Vision and Pattern Recognition Workshops}, 2019, pp. 0--0.

\bibitem{yang2018low}
Q.~Yang, P.~Yan, Y.~Zhang, H.~Yu, Y.~Shi, X.~Mou, M.~K. Kalra, Y.~Zhang, L.~Sun, and G.~Wang, ``Low-dose ct image denoising using a generative adversarial network with wasserstein distance and perceptual loss,'' \emph{IEEE transactions on medical imaging}, vol.~37, no.~6, pp. 1348--1357, 2018.

\bibitem{luo2021gpr}
J.~Luo, W.~Lei, F.~Hou, C.~Wang, Q.~Ren, S.~Zhang, S.~Luo, Y.~Wang, and L.~Xu, ``Gpr b-scan image denoising via multi-scale convolutional autoencoder with data augmentation,'' \emph{Electronics}, vol.~10, no.~11, p. 1269, 2021.

\bibitem{tran2020gan}
L.~D. Tran, S.~M. Nguyen, and M.~Arai, ``Gan-based noise model for denoising real images,'' in \emph{Proceedings of the Asian Conference on Computer Vision}, 2020.

\bibitem{hong2020end}
Z.~Hong, X.~Fan, T.~Jiang, and J.~Feng, ``End-to-end unpaired image denoising with conditional adversarial networks,'' in \emph{Proceedings of the AAAI conference on artificial intelligence}, vol.~34, no.~04, 2020, pp. 4140--4149.

\bibitem{fuentes2022mid3a}
F.~Fuentes-Hurtado, T.~Delaire, F.~Levet, J.-B. Sibarita, and V.~Viasnoff, ``Mid3a: microscopy image denoising meets differentiable data augmentation,'' in \emph{2022 International Joint Conference on Neural Networks (IJCNN)}.\hskip 1em plus 0.5em minus 0.4em\relax IEEE, 2022, pp. 1--9.

\bibitem{ioffe2015batch}
S.~Ioffe and C.~Szegedy, ``Batch normalization: Accelerating deep network training by reducing internal covariate shift,'' in \emph{International conference on machine learning}.\hskip 1em plus 0.5em minus 0.4em\relax pmlr, 2015, pp. 448--456.

\bibitem{tian2020image}
C.~Tian, Y.~Xu, and W.~Zuo, ``Image denoising using deep cnn with batch renormalization,'' \emph{Neural Networks}, vol. 121, pp. 461--473, 2020.

\bibitem{kim2020transfer}
Y.~Kim, J.~W. Soh, G.~Y. Park, and N.~I. Cho, ``Transfer learning from synthetic to real-noise denoising with adaptive instance normalization,'' in \emph{Proceedings of the IEEE/CVF Conference on Computer Vision and Pattern Recognition}, 2020, pp. 3482--3492.

\bibitem{allen1971mean}
D.~M. Allen, ``Mean square error of prediction as a criterion for selecting variables,'' \emph{Technometrics}, vol.~13, no.~3, pp. 469--475, 1971.

\bibitem{kingma2014adam}
D.~P. Kingma and J.~Ba, ``Adam: A method for stochastic optimization,'' \emph{arXiv preprint arXiv:1412.6980}, 2014.

\bibitem{ba2016layer}
J.~L. Ba, J.~R. Kiros, and G.~E. Hinton, ``Layer normalization,'' \emph{arXiv preprint arXiv:1607.06450}, 2016.

\bibitem{huang2015single}
J.-B. Huang, A.~Singh, and N.~Ahuja, ``Single image super-resolution from transformed self-exemplars,'' in \emph{Proceedings of the IEEE Conference on Computer Vision and Pattern Recognition}, 2015, pp. 5197--5206.

\bibitem{arbelaez2010contour}
P.~Arbelaez, M.~Maire, C.~Fowlkes, and J.~Malik, ``Contour detection and hierarchical image segmentation,'' \emph{IEEE Transactions on Pattern Analysis and Machine Intelligence}, vol.~33, no.~5, pp. 898--916, 2010.

\bibitem{agustsson2017ntire}
E.~Agustsson and R.~Timofte, ``Ntire 2017 challenge on single image super-resolution: Dataset and study,'' in \emph{Proceedings of the IEEE Conference on Computer Vision and Pattern Recognition Workshops}, 2017, pp. 126--135.

\bibitem{timofte2017ntire}
R.~Timofte, E.~Agustsson, L.~Van~Gool, M.-H. Yang, and L.~Zhang, ``Ntire 2017 challenge on single image super-resolution: Methods and results,'' in \emph{Proceedings of the IEEE Conference on Computer Vision and Pattern Recognition Workshops}, 2017, pp. 114--125.

\bibitem{ma2016waterloo}
K.~Ma, Z.~Duanmu, Q.~Wu, Z.~Wang, H.~Yong, H.~Li, and L.~Zhang, ``Waterloo exploration database: New challenges for image quality assessment models,'' \emph{IEEE Transactions on Image Processing}, vol.~26, no.~2, pp. 1004--1016, 2016.

\bibitem{abdelhamed2018high}
A.~Abdelhamed, S.~Lin, and M.~S. Brown, ``A high-quality denoising dataset for smartphone cameras,'' in \emph{Proceedings of the IEEE Conference on Computer Vision and Pattern Recognition}, 2018, pp. 1692--1700.

\bibitem{tian2024cross}
C.~Tian, M.~Zheng, W.~Zuo, S.~Zhang, Y.~Zhang, and C.-W. Lin, ``A cross transformer for image denoising,'' \emph{Information Fusion}, vol. 102, p. 102043, 2024.

\bibitem{martin2001database}
D.~Martin, C.~Fowlkes, D.~Tal, and J.~Malik, ``A database of human segmented natural images and its application to evaluating segmentation algorithms and measuring ecological statistics,'' in \emph{Proceedings Eighth IEEE International Conference on Computer Vision. ICCV 2001}, vol.~2.\hskip 1em plus 0.5em minus 0.4em\relax IEEE, 2001, pp. 416--423.

\bibitem{zhang2011color}
L.~Zhang, X.~Wu, A.~Buades, and X.~Li, ``Color demosaicking by local directional interpolation and nonlocal adaptive thresholding,'' \emph{Journal of Electronic Imaging}, vol.~20, no.~2, pp. 023\,016--023\,016, 2011.

\bibitem{franzen1999kodak}
R.~Franzen, ``Kodak lossless true color image suite,'' \emph{source: http://r0k. us/graphics/kodak}, vol.~4, no.~2, 1999.

\bibitem{hughes2014automated}
M.~J. Hughes and D.~J. Hayes, ``Automated detection of cloud and cloud shadow in single-date landsat imagery using neural networks and spatial post-processing,'' \emph{Remote Sensing}, vol.~6, no.~6, pp. 4907--4926, 2014.

\bibitem{nam2016holistic}
S.~Nam, Y.~Hwang, Y.~Matsushita, and S.~J. Kim, ``A holistic approach to cross-channel image noise modeling and its application to image denoising,'' in \emph{Proceedings of the IEEE Conference on Computer Vision and Pattern Recognition}, 2016, pp. 1683--1691.

\bibitem{tian2022heterogeneous}
C.~Tian, Y.~Zhang, W.~Zuo, C.-W. Lin, D.~Zhang, and Y.~Yuan, ``A heterogeneous group cnn for image super-resolution,'' \emph{IEEE transactions on neural networks and learning systems}, 2022.

\bibitem{yu2015multi}
F.~Yu and V.~Koltun, ``Multi-scale context aggregation by dilated convolutions,'' \emph{arXiv preprint arXiv:1511.07122}, 2015.

\bibitem{dabov2007image}
K.~Dabov, A.~Foi, V.~Katkovnik, and K.~Egiazarian, ``Image denoising by sparse 3-d transform-domain collaborative filtering,'' \emph{IEEE Transactions on Image Processing}, vol.~16, no.~8, pp. 2080--2095, 2007.

\bibitem{chen2016trainable}
Y.~Chen and T.~Pock, ``Trainable nonlinear reaction diffusion: A flexible framework for fast and effective image restoration,'' \emph{IEEE Transactions on Pattern Analysis and Machine Intelligence}, vol.~39, no.~6, pp. 1256--1272, 2016.

\bibitem{luo2015adaptive}
E.~Luo, S.~H. Chan, and T.~Q. Nguyen, ``Adaptive image denoising by targeted databases,'' \emph{IEEE Transactions on Image Processing}, vol.~24, no.~7, pp. 2167--2181, 2015.

\bibitem{zhang2017learning}
K.~Zhang, W.~Zuo, S.~Gu, and L.~Zhang, ``Learning deep cnn denoiser prior for image restoration,'' in \emph{Proceedings of the IEEE Conference on Computer Vision and Pattern Recognition}, 2017, pp. 3929--3938.

\bibitem{zhang2018ffdnet}
K.~Zhang, W.~Zuo, and L.~Zhang, ``Ffdnet: Toward a fast and flexible solution for cnn-based image denoising,'' \emph{IEEE Transactions on Image Processing}, vol.~27, no.~9, pp. 4608--4622, 2018.

\bibitem{plotz2018neural}
T.~Pl{\"o}tz and S.~Roth, ``Neural nearest neighbors networks,'' \emph{Advances in Neural Information Processing Systems}, vol.~31, 2018.

\bibitem{jia2019focnet}
X.~Jia, S.~Liu, X.~Feng, and L.~Zhang, ``Focnet: A fractional optimal control network for image denoising,'' in \emph{Proceedings of the IEEE/CVF Conference on Computer Vision and Pattern Recognition}, 2019, pp. 6054--6063.

\bibitem{zhang2021plug}
K.~Zhang, Y.~Li, W.~Zuo, L.~Zhang, L.~Van~Gool, and R.~Timofte, ``Plug-and-play image restoration with deep denoiser prior,'' \emph{IEEE Transactions on Pattern Analysis and Machine Intelligence}, vol.~44, no.~10, pp. 6360--6376, 2021.

\bibitem{mou2021dynamic}
C.~Mou, J.~Zhang, and Z.~Wu, ``Dynamic attentive graph learning for image restoration,'' in \emph{Proceedings of the IEEE/CVF International Conference on Computer Vision}, 2021, pp. 4328--4337.

\bibitem{zhang2023robust}
Q.~Zhang, J.~Xiao, C.~Tian, J.~Chun-Wei~Lin, and S.~Zhang, ``A robust deformed convolutional neural network (cnn) for image denoising,'' \emph{CAAI Transactions on Intelligence Technology}, vol.~8, no.~2, pp. 331--342, 2023.

\bibitem{yin2022csformer}
H.~Yin and S.~Ma, ``Csformer: Cross-scale features fusion based transformer for image denoising,'' \emph{IEEE Signal Processing Letters}, vol.~29, pp. 1809--1813, 2022.

\bibitem{zamir2022restormer}
S.~W. Zamir, A.~Arora, S.~Khan, M.~Hayat, F.~S. Khan, and M.-H. Yang, ``Restormer: Efficient transformer for high-resolution image restoration,'' in \emph{Proceedings of the IEEE/CVF Conference on Computer Vision and Pattern Recognition}, 2022, pp. 5728--5739.

\bibitem{li2021efficient}
W.~Li, X.~Lu, J.~Lu, X.~Zhang, and J.~Jia, ``On efficient transformer and image pre-training for low-level vision,'' \emph{arXiv preprint arXiv:2112.10175}, 2021.

\bibitem{hore2010image}
A.~Hore and D.~Ziou, ``Image quality metrics: Psnr vs. ssim,'' in \emph{2010 20th international conference on pattern recognition}.\hskip 1em plus 0.5em minus 0.4em\relax IEEE, 2010, pp. 2366--2369.

\bibitem{zhang2011fsim}
L.~Zhang, L.~Zhang, X.~Mou, and D.~Zhang, ``Fsim: A feature similarity index for image quality assessment,'' \emph{IEEE transactions on Image Processing}, vol.~20, no.~8, pp. 2378--2386, 2011.

\bibitem{zhang2018unreasonable}
R.~Zhang, P.~Isola, A.~A. Efros, E.~Shechtman, and O.~Wang, ``The unreasonable effectiveness of deep features as a perceptual metric,'' in \emph{Proceedings of the IEEE Conference on Computer Vision and Pattern Recognition}, 2018, pp. 586--595.

\bibitem{russo2015new}
F.~Russo \emph{et~al.}, ``New method for measuring the detail preservation of noise removal techniques in digital images,'' \emph{WSEAS Transactions on Signal Processing}, vol.~11, pp. 317--327, 2015.

\bibitem{sharma2005ciede2000}
G.~Sharma, W.~Wu, and E.~N. Dalal, ``The ciede2000 color-difference formula: Implementation notes, supplementary test data, and mathematical observations,'' \emph{Color Research \& Application: Endorsed by Inter-Society Color Council, The Colour Group (Great Britain), Canadian Society for Color, Color Science Association of Japan, Dutch Society for the Study of Color, The Swedish Colour Centre Foundation, Colour Society of Australia, Centre Fran{\c{c}}ais de la Couleur}, vol.~30, no.~1, pp. 21--30, 2005.

\bibitem{anwar2019real}
S.~Anwar and N.~Barnes, ``Real image denoising with feature attention,'' in \emph{Proceedings of the IEEE/CVF International Conference on Computer Vision}, 2019, pp. 3155--3164.

\bibitem{yue2019variational}
Z.~Yue, H.~Yong, Q.~Zhao, D.~Meng, and L.~Zhang, ``Variational denoising network: Toward blind noise modeling and removal,'' \emph{Advances in Neural Information Processing Systems}, vol.~32, 2019.

\bibitem{chang2020spatial}
M.~Chang, Q.~Li, H.~Feng, and Z.~Xu, ``Spatial-adaptive network for single image denoising,'' in \emph{Computer Vision--ECCV 2020: 16th European Conference, Glasgow, UK, August 23--28, 2020, Proceedings, Part XXX 16}.\hskip 1em plus 0.5em minus 0.4em\relax Springer, 2020, pp. 171--187.

\bibitem{yue2020dual}
Z.~Yue, Q.~Zhao, L.~Zhang, and D.~Meng, ``Dual adversarial network: Toward real-world noise removal and noise generation,'' in \emph{Computer Vision--ECCV 2020: 16th European Conference, Glasgow, UK, August 23--28, 2020, Proceedings, Part X 16}.\hskip 1em plus 0.5em minus 0.4em\relax Springer, 2020, pp. 41--58.

\bibitem{ren2021adaptive}
C.~Ren, X.~He, C.~Wang, and Z.~Zhao, ``Adaptive consistency prior based deep network for image denoising,'' in \emph{Proceedings of the IEEE/CVF Conference on Computer Vision and Pattern Recognition}, 2021, pp. 8596--8606.

\bibitem{zamir2020cycleisp}
S.~W. Zamir, A.~Arora, S.~Khan, M.~Hayat, F.~S. Khan, M.-H. Yang, and L.~Shao, ``Cycleisp: Real image restoration via improved data synthesis,'' in \emph{Proceedings of the IEEE/CVF conference on computer vision and pattern recognition}, 2020, pp. 2696--2705.

\bibitem{multiZamir}
{Zamir, Syed Waqas and Arora, Aditya and Khan, Salman and Hayat, Munawar and Khan, Fahad Shahbaz and Yang, Ming-Hsuan and Shao, Ling}, ``Multi-stage progressive image restoration,'' in \emph{Proceedings of the IEEE/CVF conference on computer vision and pattern recognition}, 2021, pp. 14\,821--14\,831.

\bibitem{ji2021u2}
H.~Ji, X.~Feng, W.~Pei, J.~Li, and G.~Lu, ``U2-former: A nested u-shaped transformer for image restoration,'' \emph{arXiv preprint arXiv:2112.02279}, 2021.

\bibitem{zhao2023comprehensive}
H.~Zhao, Y.~Gou, B.~Li, D.~Peng, J.~Lv, and X.~Peng, ``Comprehensive and delicate: An efficient transformer for image restoration,'' in \emph{Proceedings of the IEEE/CVF conference on computer vision and pattern recognition}, 2023, pp. 14\,122--14\,132.

\end{thebibliography}

\end{document}